# Metal oxide nanomaterials for pseudocapacitors

Dr. Jing Zhang[1], Dr. Yingxue Cui[1], and Prof. Dr. Guangcun Shan[2,3]*

[1]Department of Energy and Materials Engineering, Dongguk University-Seoul, Seoul 04620, Republic of Korea

[2]School of Instrument Science and Opto-electronic Engineering, Beihang University, Beijing 100083, China

[3]Department of Materials Science & Engineering, City University of Hong Kong, Hong Kong SAR, China

## 1. Introduction

With the rapid development of economy, the consumption of fossil fuels, and the increasing environment pollution, there is in urgent need of seeking clean and renewable energy sources, as well as highly efficient and low-cost energy storage technologies (Yang et al. 2011; Xiang et al. 2012; Choi et al. 2012; Dreyer et al. 2010; Wang, Zhang, et al. 2012). With the advances of technology, some kind of useful technologies for energy storage have been studied, such as batteries, fuel cells, electrochemical capacitors (ECs), and so on (Hollenkamp et al. 2009; Tasnin and Saikia 2018). **Figure 1** shows the Ragone plot comparison, supercapacitors (SCs), also called ECs, hold a significant position because of the superior energy density than traditional electrostatic capacitors and the faster power delivery than batteries (Ali et al. 2017; Jayalakshmi and Balasubramanian 2008; Melot and Tarascon 2013; Cao and Wei 2014).

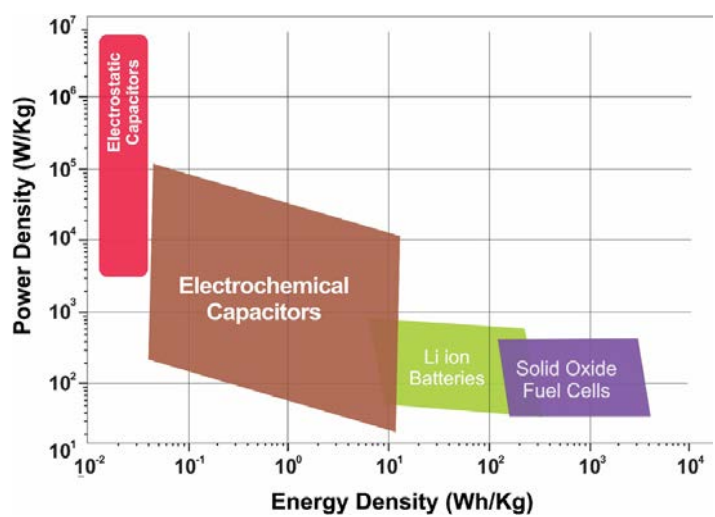

**Figure 1.** Ragone plot of different energy storage devices. Reproduced with permission (Ali et al. 2017). Copyright 2017, AIP Publishing.

In 1957, the first patent about SCs was submitted. However, not until the 1990s, did researchers start to pay attention to the application potential of SCs in the field of hybrid electric vehicles (Kötz and Carlen 2000). There are many factors affecting the electrochemical performance of SCs, including the physicochemical properties of electrode materials, the selection of electrolyte, and the voltage range of electrodes (Zhang and Pan 2015). At present, to increase the electrochemical performance of SCs, one of the most useful ways is the development of novel electrode materials with appropriate compositions and structures for accelerating electron transfer and ion diffusion (Deka et al. 2017; Wang, Wu, et al. 2017). **Figure 2** displays the yearly development of novel electrode materials for boosting the electrochemical performance of SCs (Raza et al. 2018). It is found that the most popular electrode materials for SCs today are carbon materials with high surface areas, which store charges by physical adsorption of ions on the material surface (Simon et al. 2014). The SCs of this kind of charge storage mechanism have a low energy density (<10 Wh $kg^{-1}$). In order to enhance the energy density of SCs, a lot of work has been done, such as the replacement of carbon materials using electrochemically active materials or the hybridization of carbon materials with electrochemically active materials. Compared with carbon materials, electrochemically active materials can store more energy due to the faradaic charge transfer and the electrochemical process (Wang, Zhang, et al. 2012; Miller and Simon 2008).

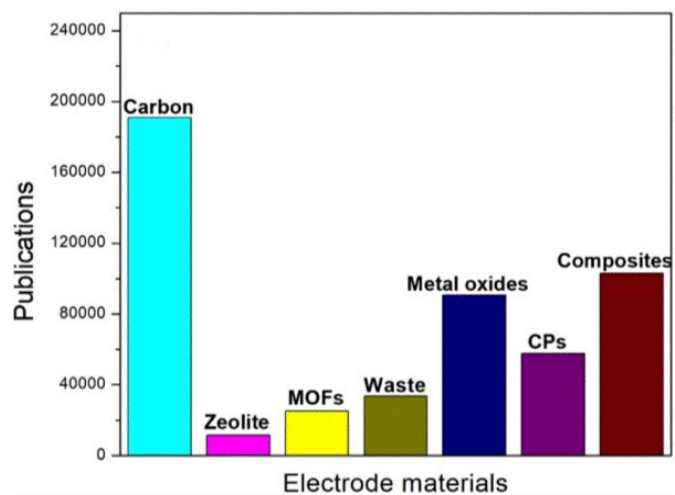

**Figure 2.** The yearly development of novel electrode materials. Reproduced with permission (Raza et al. 2018). Copyright 2018, Elsevier.

Concerning electrochemically active materials, transition metal oxides (TMOs) are regarded as the most promising candidates for the next generation SCs and have been widely reported (Wang, Zhang, et al. 2012; Mahmood et al. 2018; Guo et al. 2018; Augustyn et al. 2014; Ellis et al. 2014; Devan et al. 2012). Therefore, in this review, we mainly pay attention to TMOs with excellent electrochemical performance. This review is divided into four parts: (a) the background including energy storage mechanisms and electrochemical characterizations of SCs; (b) various kinds of TMOs for SCs are analyzed in detail involved in crystal structure, conductivity, and energy storage mechanism; (c) hybridizations with other materials to improve the electrochemical performance of TMOs are summarized and discussed; (d) a perspective and short remarks about the development and design of TMOs for superior performance SCs are shown.

## 2. Background

### 2.1 Energy storage mechanisms of SCs

As shown in **Figure 3**, the energy storage mechanisms of SCs can be divided into three types: (a) electrical double layered capacitance (EDLC), (b) pseudocapacitance, and (c) battery-type behavior (Wang et al. 2018). More details about these three energy storage mechanisms and related electrode materials are presented in the following.

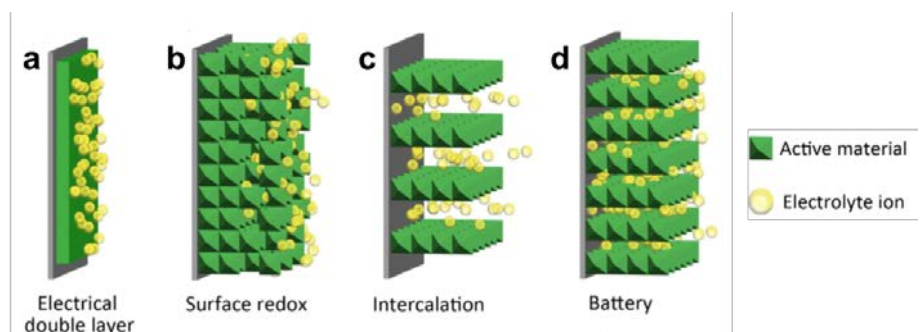

**Figure 3.** Energy storage mechanism illustration of SCs: (a) electrical double layer capacitance, (b) surface redox capacitance, (c) intercalation capacitance, and (d) battery-type behavior. Reproduced with permission (Wang et al. 2018). Copyright 2019, Elsevier.

**2.1.1** *Mechanism of EDLC*

The energy storage mechanism of EDLC is based on the absorption/desorption of electrolyte ions on the electrode/electrolyte interface during the charge-discharge process without the charge transfer reaction, as shown in **Figure 3a** (Wang et al. 2018; Xu, Shi, et al. 2015; Zhang et al. 2010). The EDLC behavior in the purely physical process largely relies on the electrochemical activity and kinetic feature of the electrode materials for the absorption/desorption of electrolyte ions (Meng et al. 2017). Therefore, to receive outstanding capacitance, it is needed to optimize the surface area, porosity and pore distribution of the electrode materials (Itoi et al. 2011). Largeot et al. (Largeot et al. 2008) reported the relationship between the electrolyte ion size and the pore size of the electrode material in an EDLC. And it is found that the EDLC can receive the highest double-layer capacitance when the electrolyte ion size is close to the pore size of the electrode material.

Carbon materials (such as graphene, carbon nanotubes (CNTs), active carbon, and so on) with large surface area, porous structures, and high conductivity, show EDLC behaviors (Xu, Shi, et al. 2015; Simon and Gogotsi 2012; Yang, Ren, et al. 2015, Wang, Ren, Shan, et al. 2019). However, the specific capacitances of carbon materials (<200 F $g^{-1}$).are lower than that of faradaic active materials.

**2.1.2** *Mechanism of pseudocapacitance*

Unlike the EDLC that physically accumulates charges, pseudocapacitance endows charge storage capacitance by faradaic redox reactions. Fundamentally, pseudocapacitance can be identified as two types, which is illustrated in **Figure 3b,c**: surface redox pseudocapacitance and intercalation pseudocapacitance (Augustyn et al. 2014; Wang et al. 2018).

Surface redox pseudocapacitance occurs at surface or near-surface of electrode material when electrolyte cations and anions are adsorbed/desorbed, accompanied by charge storage process (Kim, Leverage, et al. 2014). Like EDLC based materials, electrode materials with surface redox pseudocapacitance have linear and triangle shape of galvanic charge/discharge (GCD) curves and nearly rectangular cyclic voltammetry (CV) curves. Recently, various electrode materials have been widely studied as the promising candidates to present surface redox pseudocapacitance, such as conducting polymers (Snook et al. 2011; Shi et al. 2015) and some TMOs ($RuO_2$ (Trasatti and Buzzanca 1971), Mn (Wei et al. 2011; Wang, Kang, et al. 2015; Hu et al. 2018) and Fe-based oxides (Zeng, Yu, et al. 2016)). For conducting polymers, the material structures are subject to change because of the volume expansion during charge/discharge process, leading to poor cyclic stability, which limits their practical SC applications (Snook et al. 2011; Shi et al. 2015). For $RuO_2$, it was the first reported transition metal oxide for SCs (Trasatti and Buzzanca 1971). The maximum theoretical capacitance of $RuO_2$ can reach 1450 F $g^{-1}$ in a potential window of 1 V, which is much higher than that of carbon-based materials. However, the high cost of Ru limits the large-scale SC application of $RuO_2$. Therefore, exploiting alternative low-cost transition metal based electroactive materials is necessary.

Intercalation pseudocapacitance involves a process that electrolyte cations ($Na^+$, $K^+$, $Li^+$, $H^+$, etc.) intercalate/deintercalate into/from the channels or layers of electrode materials accompanied by a faradaic charge-transfer involving no phase change (Yu, Yun, et al. 2018). Intercalation dominated pseudocapacitance electrode materials involve rigid layered lattice framework, including $V_2O_5$ (Liu, Su, et al. 2017; Wu et al. 2013), $WO_3$ (Zhu et al. 2014), $MoO_3$ (Hu, Zhang, et al. 2015), $Nb_2O_5$ (Yan, Rui, et al. 2016), and $TiO_2$ (Dylla et al. 2012). TMOs without two-dimensional pathway are not belong to intercalation pseudocapacitance materials. For the convenience of discussion,

$VO_2$, $V_2O_3$, and $MoO_2$ without layered lattice framework are also discussed in the corresponding part of TMOs.

**2.1.3** *Mechanism of battery-type behavior*

Electrode materials that undergo a phase change during charge/discharge process have long been mishandled as pseudocapacitance materials. Recently, the electrochemical community has distinguished above electrode materials from traditional pseudocapacitance materials. On the basis of the new definition, electrode materials exhibiting a phase change during charge/discharge process should be classified as "battery-type" materials (Simon et al. 2014; Brousse et al. 2015).[17,147] As shown in **Figure 3d**, the charge storage of battery-type materials is controlled by cation diffusion and involves ion intercalation/deintercalation reaction, phase change, and/or alloying reactions (Wang et al. 2016). The electrochemical characteristics of battery-type materials are expressed in the cyclic voltammetry (CV) and galvanostatic charge/discharge (GCD) curves with distinct redox CV peaks and GCD platform regions, respectively. Generally, battery-type materials have high specific capacity but poor rate performance, which is related to the slow kinetics associated with sluggish phase change in the charge /discharge process. However, the batter-type materials with specific morphologies can have a high specific surface area, which can provide rich active sites for redox reactions and shorten the diffusion distance of electrolyte ions. The porous nanostructures also offer adequate spaces to buffer volume expansion during charge/discharge process, which improve the cycling stability of battery-type materials. In order to improve the charge storage rate of battery-type materials, they are usually combined with carbon-based materials to form hybrid supercapacitors, thereby achieving excellent rate performance and high energy density. Ni (Feng et al. 2014; Li, Zheng, et al. 2016; Sk et al. 2016) and Co-based (Uke et al. 2017; Chang et al. 2015) oxides are typical battery-type electrode materials. They have a voltage window of 0-0.55 V vs. Hg/HgO in alkaline electrolytes. When they are assembled into hybrid SC devices with carbon-based materials (-1-0 V vs. Hg/HgO), the devices can achieve a voltage window of 1.5V. Besides Ni and Co-based oxides, other transition metal-based materials such as Cu (Zhang et al. 2014) and some binary TMOs (Chen et al. 2015; Gao et al. 2017) also exhibit battery-type charge storage behavior. These materials for SCs are discussed in detail in Section 3.3.

2.2 Electrochemical characterizations of SCs

The electrochemical performance of SCs can be defined by a number of parameters such as specific capacitance ($C_s$), voltage window ($V$), equivalent series resistance (ESR), power density ($P$), energy density ($E$), and time constant ($t$) (Zhang and Pan 2015). Generally, three techniques are used to evaluate the electrochemical performance of SCs including cyclic voltammetry (CV), galvanostatic charge/discharge (GCD), and electrochemical impedance spectroscopy (EIS).

*2.2.1 Cyclic voltammetry*

CV test can be used to evaluate the charge storage mechanism and calculate the specific capacitance of electrode materials in a three-electrode system. The three-electrode system is composed of working electrode (WE), counter electrode (CE), and reference electrode (RE), as shown in **Figure 4a**. The CV curves are obtained by their responding to voltage sweeps. The abscissa and ordinate of CV curves are potential and current density, respectively. For EDLCs, their CV curve shapes are almost rectangular due to the highly reversible adsorption/desorption of electrolyte ions at the electrode/electrolyte interface (Yang, Ren, et al. 2015). For pseudocapacitors, their CV curve shapes are rectangular or redox peaks due to the redox reactions of electrode materials (Kim, Leverage, et al. 2014; Yu, Yun, et al. 2018; Brousse et al. 2015). The specific capacitance of the electrode materials obtained by the CV curves can be calculated according to the following **equation (1)** (Conway 2013):

$$C_s = \frac{\int I dV}{v \Delta V m} \tag{1}$$

where $C_s$, $I$ (A), $V$ (V), $v$ (V s$^{-1}$), $\Delta V$ (V), and $m$ (g), stand for the specific capacitance, the instantaneous current, the potential, the potential scan rate, the voltage change, the mass of the active material, and the discharge time, respectively.

### 2.2.2 Galvanostatic charge/discharge

GCD curves are obtained by their response to constant current and can also characterize capacitive behaviors. For EDLCs, their GCD curves have linear and triangle shape (Yang, Ren, et al. 2015). For pseudocapacitors, their GCD curves are similar to that of EDLC based materials or have GCD platform regions (Kim, Leverage, et al. 2014; Yu, Yun, et al. 2018; Brousse et al. 2015). In a three-electrode system, the specific capacitance of electrode materials can be calculated from the GCD plots according to the following **equation (2)** (Conway 2013):

$$C_s = \frac{I \Delta t}{m \Delta V} \tag{2}$$

where $C_s$, $I$, $\Delta t$, $m$, $\Delta V$ represent the specific capacitance, the discharge current, the discharge time, the mass of the active material, and the voltage change.

Energy density and power density are also two important parameters for evaluating the performance of energy storage devices. SC device is a two-electrode system composed of a pair of closely spaced electrodes including the active materials attached to the current collector saturated with electrolyte and a separator sandwiched between the parallel electrodes, as shown in **Figure 4b**. Energy density represents the total amount of charge stored in the SCs per unit mass or volume, and power density is related to the rate of charge delivered during discharge. Energy density and power density are expressed by the following **equation (3)** and **(4)** (Conway 2013):

$$E = \frac{C_s \Delta V}{2} \times \frac{1000}{3600} \tag{3}$$

$$P = \frac{E}{\Delta t} \times 3600 \tag{4}$$

where $E$ (Wh kg$^{-1}$), $P$ (W kg$^{-1}$), $C_s$ (F g$^{-1}$), $\Delta V$ (V), and $\Delta t$ (s) represent the energy density, the power density, the specific capacitance, the maximum voltage, and the discharge time, respectively.

### 2.2.3 Electrochemical impedance spectroscopy

EIS is used to study the transport characteristics of charge carriers in SCs. Generally, for EIS tests, the impedance data are collected at an open circuit potential by applying a small amplitude (e.g., ±5 to ±10 mV) of alternating potential over a wide frequency range (e.g., from 0.01 to 100 kHz), which are usually displayed in a Nyquist plot (Zhang and Zhao 2012). In the Nyquist plot, the expression of impedance (Z) can be simplified as:

$$Z = Z' + jZ'' \tag{5}$$

where Z' and Z'' represent the real and imaginary part of the impedance, respectively.

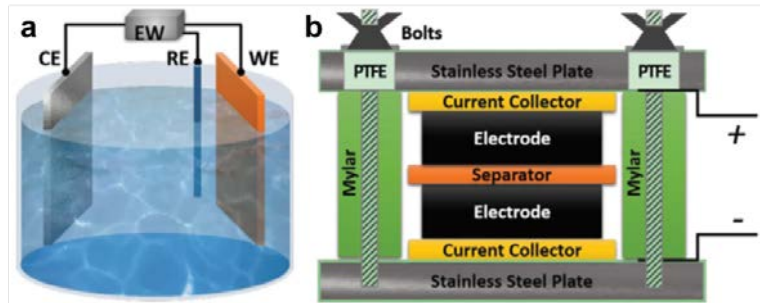

**Figure 4.** Schematic diagram of (a) three-electrode system (WE: working electrode, CE: counter electrode, and RE: reference electrode) and (b) two-electrode system. Reproduced with permission (Wu et al. 2017). Copyright 2017, Wiley.

## 3. TMOs-based electrode materials for SCs

**3.1** Surface redox reaction-based pseudocapacitive electrode materials

**3.1.1** *Ruthenium oxides*

Ruthenium oxide materials, such as $RuO_2$ and $RuO_2 \cdot xH_2O$ are typical surface redox pseudocapacitive materials. Ruthenium oxides as electrode materials for SCs have attracted a lot of attention because of their high conductivity, good thermal stability, and high thermal stability (Huang et al. 2013). The energy storage mechanism of $RuO_2$ can be described as **equation (6)** or **(7)** (Wu et al. 2017):

$$RuO_2 + xH^+ + xe^- = RuO_{2-x}(OH)_x \ (0 < x < 2) \tag{6}$$

or

$$RuO_2 + H^+ + e^- = RuOOH \tag{7}$$

The phase of crystalline ruthenium oxides plays an important role in deciding their electrochemical performance. Crystalline ruthenium oxides have two different phases, including rutile phase $RuO_2$ and hydrated $RuO_2 \cdot xH_2O$ (Wu et al. 2017).[33] Sugimoto et al. (Sugimoto et al. 2005) compared the electrochemical performance of these two phases. As shown in **Figure 5 a,b**, anhydrous $RuO_2$ displays a specific capacitance of 24 F $g^{-1}$ in a wide voltage range (1.0 V), while hydrated $RuO_2 \cdot 0.5H_2O$ possesses a higher specific capacitance of 342 F $g^{-1}$. The pseudocapacitance of $RuO_2 \cdot 0.5H_2O$ can be derived from the easily accessible mesopores (thick trunk roots in **Figure 4c**) and the less accessible micropores (thin roots in Figure 4c). However, most of pseudocapacitance in anhydrous $RuO_2$ only comes from the easily accessible mesopores (thick trunk roots in **Figure 5d**). The utilization of micropores determines the capacitance. Therefore, the specific capacitance of hydrated $RuO_2$ is higher than that of anhydrous $RuO_2$. Compared with crystalline $RuO_2$, amorphous $RuO_2$ presented a higher specific capacitance of 1580 F $g^{-1}$, because the flexibility of amorphous structure can achieve lattice rearrangement based on ion intercalation/deintercalation, leading to more active sites for redox reaction (Gujar et al. 2007).

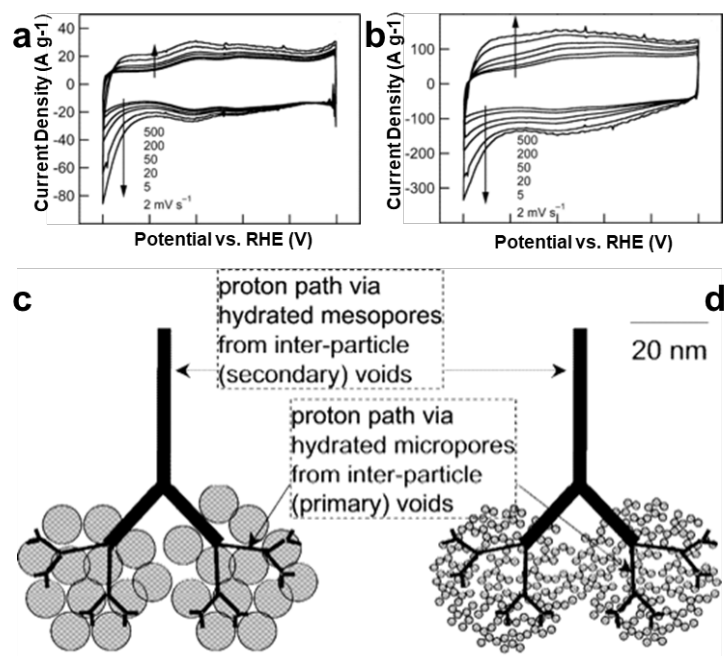

**Figure 5.** (a,b) CV curves at various scan rates and (c,d) schematic diagrams of fractal tree-root model of (a,c) anhydrous $RuO_2$ and (b,d) hydrous $RuO_2$. Reproduced with permission (Sugimoto et al. 2005). Copyright 2005, American Chemical Society.

The major issue for ruthenium oxide materials is their easy agglomerations, which severely reduces their electrochemical performance. Hence, in order to increase the pseudocapacitance of ruthenium oxide-based electrode, the good dispersion of ruthenium oxides becomes important (Ardizzone et al. 1990). Hybridization with carbon-based materials was constructed to improve the dispersion of ruthenium oxides (Jiang et al. 2012). For example, $RuO_2 \cdot xH_2O$ nanoparticles were highly dispersed on active carbon and exhibited a high specific capacitance of 1340 F $g^{-1}$ (Hu et al. 2004). Other carbon materials, such as Ketjen Black (Naoi et al. 2009), CNTs (Kim et al. 2005), and graphene (Wu, Wang, et al. 2010), can also be used as supports for the dispersion of ruthenium oxides.

Although ruthenium oxides are ideal electrode materials for high-performance SCs, it is prohibitive for large-scale commercial applications because of their scarcity and high cost. Therefore, it is urgent to dope non-noble metals into ruthenium oxides or find low-cost alternative electrode materials for the development of SCs.

### 3.1.2 *Manganese oxides*

Manganese oxides ($MnO_2$ and $Mn_3O_4$) as pseudocapacitive electrode materials have received a lot of attention due to their high specific capacitance, low price, abundant reserve, and environmental friendliness since Lee and Goodenough proposed their pseudocapacitive behaviors in 1999 (Lee and Goodenough 1999; Lee et al. 1999; Li et al. 2018). In general, the charge storage of $MnO_2$ is based on the change of manganese oxidation state from +3 to +4 and the participation of protons or alkali cations, which can be described as **equation (8)** (Zhang, Han, et al. 2015):

$$MnO_2 + C^+ + e^- \leftrightarrow MnOOC \tag{8}$$

where $C^+$ represents protons or alkali cations ($Li^+$, $Na^+$, $K^+$).

Crystal structure can affect the pseudocapacitance of $MnO_2$ electrodes (Wei et al. 2011). $MnO_2$ materials have several crystalline structures including $α$, $β$, $γ$, $δ$, and $λ$ on the basis of the interlinked ways of $MnO_6$ octahedra, as shown in **Figure 6** (Devaraj and Munichandraiah 2008). Among them, $α$, $β$, and $γ$-$MnO_2$ have 1D tunnel structures (2 × 2 octahedral units for $α$-$MnO_2$ and 1 × 1 octahedral units for $β$-$MnO_2$), $δ$-$MnO_2$ has a 2D layer structure, and $λ$-$MnO_2$ has a 3D spinel structure. The electrode materials with sufficient porosity can accelerate the insertion/deinsertion of electrolyte ions, leading to a higher specific capacitance (Huang, Li, et al. 2015). Therefore, diverse polymorph $MnO_2$ with different tunnel sizes have a significant influence on cation transport and insertion (Wei et al. 2011). $α$-$MnO_2$ with channel size of 4.6 Å and $δ$-$MnO_2$ with an interlayer spacing of 7.0 Å can accommodate $K^+$ with an ionic diameter of 3.0 Å and smaller ions to intercalate/deintercalate. However, $λ$-$MnO_2$ and $β$-$MnO_2$ with small channel size (<3.0 Å) can impede the insertion and deinsertion of $K^+$, which results in a lower specific capacitance (Huang, Li, et al. 2015).

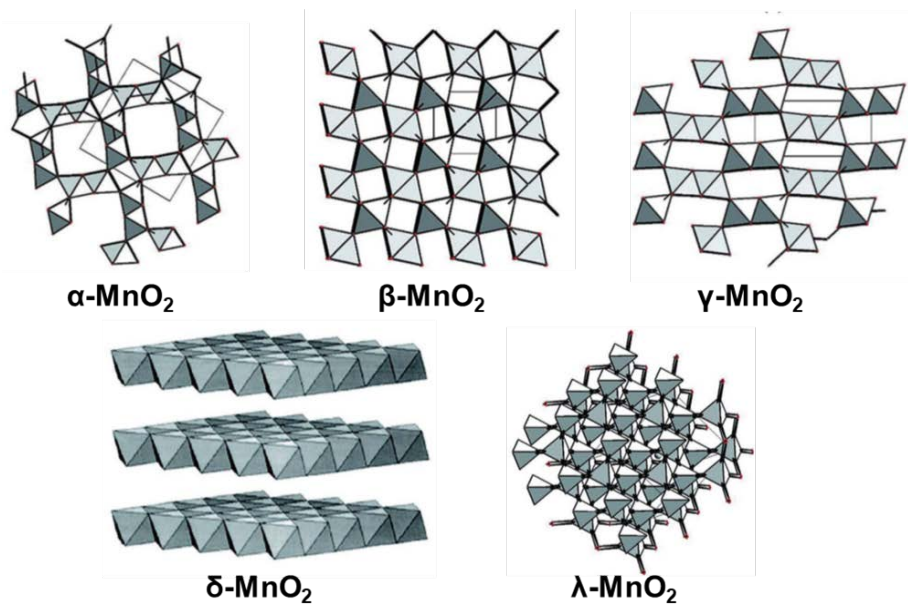

**Figure 6.** Crystal structures of α, β, γ, δ, and λ-$MnO_2$. Reproduced with permission (Devaraj and Munichandraiah 2008). Copyright 2008, American Chemical Society.

The main problems of $MnO_2$-based materials include poor electroconductivity (∼$10^{-5}$-$10^{-6}$ S $cm^{-1}$), high dissolubility in alkaline or neutral electrolyte, and specific surface area, which can hinder their electron transport rate, cycle life, and specific surface area (Wang, Zhang, et al. 2012). To improve the electrical conductivity of $MnO_2$-based materials, doping with Ru (Kim, Kim, et al. 2014), Ni (Chen and Hu 2003), and Mo (Nakayama et al. 2005) have been carried out. The reassembled Li-$Mn_{1-x}Ru_xO_2$ presented outstanding pseudocapacitive property with a high specific capacitance of 360 F $g^{-1}$, which was higher than that of unsubstituted Li-$MnO_2$ and many other $MnO_2$-based materials (Kim, Kim, et al. 2014). Besides this method, some supports with high conductivity such as carbon materials (active carbon, CNTs, and graphite) and conducting polymers have also been used to promote the conductivity of $MnO_2$-based materials (Wang, Kang, et al. 2015; Hu et al. 2018; Zhang, Zhang, et al. 2018). Conducting polymers can also serve as protective shells to prevent $MnO_2$-based materials from dissolving during cycling (Babakhani and Ivey 2010; Sharma et al. 2008; Komaba et al. 2008). The morphology can influence the specific surface area and thus the specific capacitance. Therefore, to enhance the specific surface areas of $MnO_2$-based materials, their morphologies can be controlled from 0D to 3D (zero-dimensional, one-dimensional, two-dimensional, and three-dimensional) (Hu et al. 2018). Based on the morphologies, the specific surface areas of obtained materials can range from 20 to 306.9 $m^2\ g^{-1}$.

0D nanostructures are spherical particles, including solid, hollow, and core-shell nanostructures. $MnO_2$-based 0D solid nanoparticles have been reported for SCs (Kiani et al. 2014; Jian et al. 2017). Jian et al. (Jian et al. 2017) decorated $MnO_2$ nanoparticles on 3D graphite-like capsules (GCs) using the microwave method, as shown in **Figure 7a**. The specific surface area of $MnO_2@GCs@MnO_2$ can reach 190 $m^2\ g^{-1}$, which can bring abundant active sites and space for energy storage. The specific capacitance of $MnO_2@GCs@MnO_2$ can reach 390 $F\ g^{-1}$ at a current density of 0.5 $A\ g^{-1}$ in 6 M KOH electrolyte. For hollow 0D nanostructures, they possess not only low density and high surface-to-volume ratio but also shortened pathways for ion transport (Lai et al. 2012). Several kinds of literature about $MnO_2$ 0D hollow structures have been reported (Tang et al. 2009; Bian et al. 2013). Tang et al. (Tang et al. 2009) reported the synthesis of hollow $MnO_2$ nanospheres by templating-assisted hydrothermal method. The obtained hollow $MnO_2$ nanospheres displayed a high specific surface area of 253 $m^2\ g^{-1}$. With such a structure design, a high specific capacitance of 299 $F\ g^{-1}$ at 5 $mV\ s^{-1}$ can obtain. Core-shell 0D nanostructures are composed of a solid or hollow core and a thin shell. Electrode materials composed of faradaic and non-faradaic materials with core-shell 0D nanostructures can provide many advantages including large specific surface area, outstanding conductivity, good dispersibility, robust chemical and mechanical stability (Bian et al. 2013; Zhou et al. 2017; Yu et al. 2017). For example, Lei et al. (Lei et al.

2012) prepared 0D core-shell nanostructures with $MnO_2$ shell and carbon core. The obtained C-$MnO_2$ core-shell nanostructures with a large specific surface area of 270 $m^2$ $g^{-1}$ presented a specific capacitance of 190 F $g^{-1}$ at 0.1 A $g^{-1}$. Moreover, the C-$MnO_2$ core-shell nanostructures displayed a high rate capability with 55% capacitance retention from 0.1 to 10 A $g^{-1}$, indicating porous $MnO_2$ shell can offer efficient diffusion pathways for ions.

Diverse 1D $MnO_2$ nanostructures, such as nanorods, nanowires, and nanotubes, for SCs have been reported (Zhai et al. 2014; Yao et al. 2014; Xia et al. 2010). Oxygen-deficient $MnO_2$ nanorods were synthesized by a hydrogenation treatment method and yielded a large specific capacitance of 449 F $g^{-1}$ at 0.75 mA $cm^{-2}$, as shown in **Figure 7b** (Zhai et al. 2014). The oxygen vacancy contents in $MnO_2$ lead to various capacitive performance (**Figure 7c,d**). Therefore, the optimized electrochemical activity can be achieved by adjusting the concentration of oxygen vacancies. Yao et al. (Yao et al. 2014) reported the synthesis of ultralong $MnO_2$ nanowires with a length/diameter aspect ratio of 105 using a hydrothermal method. The specific capacitance of ultralong $MnO_2$ nanowires was only 118 F $g^{-1}$ at 2 mV $s^{-1}$, which was relatively low because of the large thickness of $MnO_2$ paper. $MnO_2$ nanotube arrays were obtained by electrochemical deposition for 10 min using anodic aluminum oxide (AAO) as a template; whereas $MnO_2$ nanowire arrays were obtained after 60 min (Xia et al. 2010). A specific capacitance of 320 F $g^{-1}$ can be obtained for $MnO_2$ nanotube arrays, however, only 101 F $g^{-1}$ can be received for $MnO_2$ nanowire arrays.

2D $MnO_2$ nanostructures, such as nanosheets, nanofilms, and nanoflakes, have been extensively studied because of their large specific surface areas and abundant active sites (Cheng, Tan, et al. 2015; Lee et al. 2010; Wang and Wang 2017; Qian, Jin, et al. 2015; Liu, Wang, et al. 2017). Unlike the 0D and 1D nanostructures, 2D nanostructures can be utilized to fabricate free-standing and substrate-free electrodes. Using electrodeposition method, a $MnO_2$ nanofilm grown on active carbon paper was obtained, which showed a high specific capacitance of 640.8 F $g^{-1}$ at a current density of 10 A $g^{-1}$ (Cheng, Tan, et al. 2015). Redox-deposition method is often applied for the synthesis of MnO$_2$ nanofilm (Lee et al. 2010). However, compared to the mass loading of 2D MnO$_2$ nanofilms obtained by the hydrothermal method, those prepared by electrodeposition or redox-deposition method is lower (Wang and Wang 2017). Qian et al. (Qian, Jin, et al. 2015) designed a chemical reduction method to synthesize the 2D MnO$_2$ nanofilm, as shown in **Figure 7e**. The 2D MnO$_2$ nanofilm (Mn paper-80) with a mass loading of 0.09 mg cm$^{-2}$ can receive a high specific capacitance of 1035 F g$^{-1}$ at 2 mV s$^{-1}$ (**Figure 7f**). However, when the mass loading was elevated to 0.18 mg cm$^{-2}$, the specific capacitance of 2D MnO$_2$ nanofilm dramatically reduced to 200 F g$^{-1}$. It is noteworthy that only the surface and near the surface of a few nanometers are redox active sites for 2D MnO$_2$ nanostructures (Liu, Wang, et al. 2017). Therefore, to obtain a high specific capacitance at a high mass loading, rational architecture design for 2D MnO$_2$ nanostructures is crucial. Wang et al. (Wang and Wang 2017) came up with a sacrificial template method to prepare MnO$_2$ nanosheets. The MnO$_2$ nanosheets with an exceptionally high mass loading of 4 mg cm$^{-2}$ can also obtain a relatively high specific capacitance of 516.7 F g$^{-1}$ 0.2 A g$^{-1}$, which is mainly due to their high specific surface area of 306.9 m$^2$ g$^{-1}$. This work shows that MnO$_2$ as electrode materials can surmount some of their inherent limitations by rational architecture designs.

3D porous nanostructures can provide well-defined pathways to electrolyte access, large surface area, and mechanical stability for SC electrodes (Ellis et al. 2014; Rolison et al. 2009; Bag and Raj 2016; Zhu et al. 2017). Bag et al. (Bag and Raj 2016) proposed a template-free method to prepare 3D self-branched $\alpha$-MnO$_2$@$\delta$-MnO$_2$ (**Figure 7g,h**). This 3D self-branched $\alpha$-MnO$_2$@$\delta$-MnO$_2$ had a high specific surface area of 238 m$^2$ g$^{-1}$ and appropriate pore size of 3.6 nm, which can accelerate charge transfer and ion diffusion. However, as shown in **Figure 7i**, the specific capacitance of 3D mesoporous $\delta$-MnO$_2$ was only 364 F g$^{-1}$ at 1.8 A g$^{-1}$, which is because of their disorder and large size (>1 μm). It is noteworthy that the contacts between the active sites of 3D nanostructures and the current collector are impeded due to the large size of 3D clusters., leading to reduced charge transport. Zhu et al. (Zhu et al. 2017) designed a novel 3D MnO$_2$ nanostructures composed by $\alpha$-MnO$_2$ nanowires and $\delta$-MnO$_2$ nanoflakes. Benefited from the novel structure, the MnO$_2$ electrode with a high mass loading of

3.77 mg cm$^{-2}$ can obtain a specific capacitance of 178 F g$^{-1}$ at 5 mV s$^{-1}$. However, the length of MnO$_2$ nanowires was about 5 μm, leading to limited contact between each nanowire and the current collector. Therefore, in order to maximize the contacts between the active materials and the substrates or conductive supports, the size of active materials should be optimized in the nanometer scale.

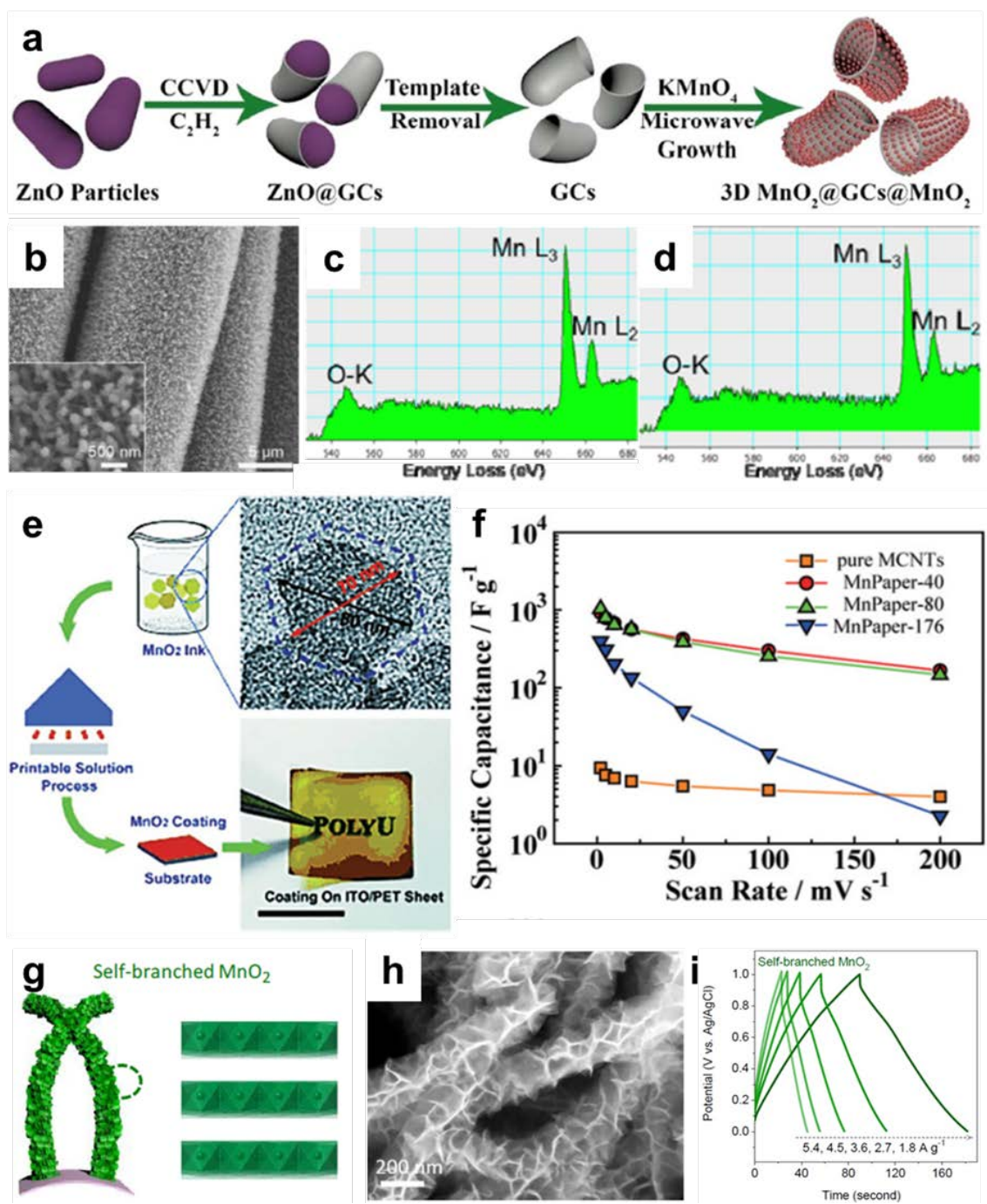

**Figure 7.** (a) Schematic illustration for the synthesis of MnO$_2$@GCs@MnO$_2$. Reproduced with

permission (Jian et al. 2017). Copyright 2017, American Chemical Society. (b) Field emission scanning electron microscopy (FE-SEM) image of oxygen-deficient $MnO_2$ nanowires, electron-energy-loss-spectra (EELS) spectra of (c) edged part and (d) central part of oxygen-deficient $MnO_2$ nanowires. Reproduced with permission (Zhai et al. 2014). Copyright 2017, Elsevier. (e) Schematic diagram of the printing process, the transmission electron microscopy (TEM) image of a single $MnO_2$ nanosheet, and the printed film on indium-tin-oxide-coated poly(ethylene terephthalate) (ITO/PET) sheet, (f) specific capacitance versus scan rate for 2D $MnO_2$ nanofilm. Reproduced with permission (Qian, Jin, et al. 2015). Copyright 2015, Wiley. (g) Schematic drawing of a self-branched $MnO_2$ electrode formed after $\delta$-$MnO_2$ flakes were decorated onto the $\alpha$-$MnO_2$ nanowires, (h) FE-SEM image, and (i) GCD curves of 3D self-branched $MnO_2$. Reproduced with permission (Zhu et al. 2017). Copyright 2017, Royal of Society of Chemistry.

$Mn_3O_4$ with spinel crystal structure can provide robust crystalline architecture with 3D diffusion channels (Lee et al. 2012; Yeager et al. 2013). Compared with $MnO_2$, $Mn_3O_4$ has poor electronic conductivity, leading to low capacitance values. The energy storage mechanism in $Mn_3O_4$ is not fully established. Recently, in situ X-ray absorption near-edge spectra (XANES) experiments have indicated that the mechanism of $Mn_3O_4$ includes the reversible redox of $Mn^{3+}$ to $Mn^{2+}$ (Yeager et al. 2013). In that study, the obtained $Mn_3O_4$ nanofilm showed the specific capacitance of 261 F $g^{-1}$.

### 3.1.3 *Iron oxides*

Iron oxides, such as $Fe_2O_3$ and $Fe_3O_4$, have received many interests as promising electrode materials for SCs because of their high theoretical capacity, wide working window in negative potential, non-toxic, earth-abundance, and low cost (Zeng, Yu, et al. 2016; Yu, Ng, et al. 2018; Xia et al. 2016; Ma et al. 2018). In order to study the energy storage mechanisms of iron oxides, a lot of efforts have been done. For example, Li et al. (Li et al. 2015) employed X-ray diffraction (XRD) and X-ray photoelectron spectroscopy (XPS) equipment to intensively study the charge mechanism of $Fe_3O_4$ in KOH electrolyte, and the results showed that a faradaic reaction occurs between $Fe^{3+}$ and $Fe^0$ in a wide voltage window. Based on the energy storage mechanism of Ni-Fe alkaline aqueous battery, the oxidation reactions of $Fe_3O_4$ can be summarized as **equation (9)** and **(10)** (Liu et al. 2014):

$Fe + 2OH^- \leftrightarrow Fe(OH)_2 + 2e^-$ ($E^0 = -1.076$ V vs. SCE) **(9)**

$$3Fe(OH)_2 + 2OH^- \leftrightarrow Fe_3O_4 + 4H_2O + 2e^- \quad (E^0 = -0.859 \text{ V vs. SCE}) \tag{10}$$

In the neutral aqueous electrolytes, one possible charge storage mechanism of $FeO_x$ is displayed as **equation (11)** (Xia et al. 2016):

$$FeO_x + yM^+ + ye^- \leftrightarrow M_yFeO_x \tag{11}$$

where M represents Li, Na, or K. Although diverse energy storage mechanisms of iron oxides have been given, the origin of pseudocapacitance and the exact redox reactions during the charge-discharge processes are still not fully established.

As is well-known, crystal forms play an important role in adjusting the electrochemical performance. For $Fe_2O_3$ materials, there are four diverse crystal structures, including $α$ (rhombohedrally centered hexagonal structure), $β$ (cubic body-centered crystal structure), $γ$ (cubic structure of inverse spinel type), and $ε$-$Fe_2O_3$ (orthorhombic structure), which are summarized in **Figure 8** (Xia et al. 2016). Compared with other crystal structures, $α$-$Fe_2O_3$ with corundum crystal structure possesses more thermodynamically stability, which can make it a promising electrode material for long life SCs (Sakurai et al. 2009). Other phases, such as $β$ and $γ$-$Fe_2O_3$, as intermediary metastable phases, can be evolved to $α$-$Fe_2O_3$ by heat treatment (Sakurai et al. 2009; Danno et al. 2013).

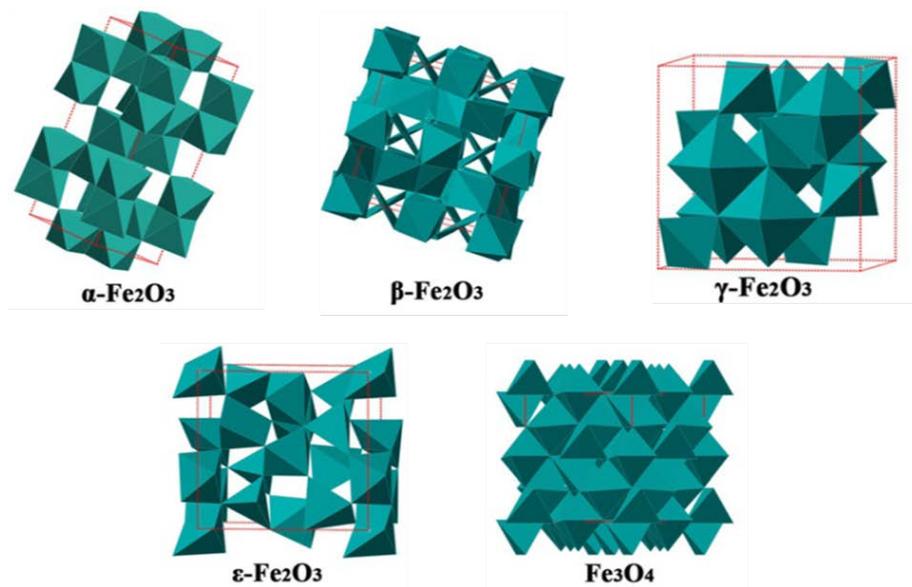

**Figure 8.** Crystal structures of $α$, $β$, $γ$, $ε$-$Fe_2O_3$, and $Fe_3O_4$. Reproduced with permission (Xia et al. 2016). Copyright 2016, Wiley.

Besides crystal structures, morphologies, compositions also have important effects on the electrochemical properties of iron oxides. Considering these factors, many methods such as nanostructure engineering and hybridization components have been proposed to improve the specific capacitance, cycle stability, and rate capacity of the iron oxides. The part of the hybridization components is detailly presented in the next section. For nanostructure engineering, this method can increase the accessible surface areas, improve the mechanical stability, and shorten the paths for electron and ion transport.

At present, various $Fe_2O_3$ nanostructures from 0D to 3D have been obtained by different methods, such as hydrothermal/solvothermal (Lu et al. 2014; Tang et al. 2015; Zheng, Yan, et al. 2016), electrospinning (Binitha et al. 2013), thermal decomposition (Lu et al. 2014; Tang et al. 2015), sol-gel (Shivakumara et al. 2014), electrodeposition (Li et al. 2017), microwave-assisted (Karthikeyan et al. 2014), template-assisted (Yang et al. 2014), and electrochemical anodization (Liu et al. 2015).

Generally, 0D nanostructures are spherical nanoparticles. For example, Shivakumara et al. (Shivakumara et al. 2014) designed $α$-$Fe_2O_3$ solid nanoparticles with a large specific area of 386 $m^2$ $g^{-1}$ using sol-gel method (**Figure 9a,b**). The obtained $α$-$Fe_2O_3$ nanoparticles (Nano $Fe_2O_3$) displayed a specific capacitance of 300 F $g^{-1}$ at 1 A $g^{-1}$ in a voltage window of 0--0.8 V (vs. Ag/AgCl) in 0.5 M $Na_2SO_3$ electrolyte (**Figure 9c**).

1D nanostructures, such as nanorods, nanowires, and nanotubes, have attracted a lot of attention due to their large specific surface area, efficient 1D electron transport pathway, and flexibility during the ion intercalation (Lu et al. 2014; Tang et al. 2015; Yang et al. 2014). Lu et al. (Lu et al. 2014) prepared 1D oxygen-deficient $α$-$Fe_2O_3$ nanorods (N-$Fe_2O_3$) on flexible carbon cloth combining solvothermal and thermal decomposition methods (**Figure 9d**). Compared with $α$-$Fe_2O_3$ without oxygen-vacancies (A-$Fe_2O_3$), the oxygen-deficient $α$-$Fe_2O_3$ nanorods (N-$Fe_2O_3$) exhibited a higher areal capacitance of 382.7 mF $cm^{-2}$ at 0.5 mA $cm^{-2}$ in a potential window of -0.8-0 V (vs. SCE) in 3 M LiCl aqueous electrolyte (**Figure 9e**). The improved pseudocapacitance of $α$-$Fe_2O_3$ nanorods can be attributed to the enhancive donor density and active sites resulting from the formation of oxygen vacancies (**Figure 9f**).

Combined solvothermal-thermal decomposition approach was also applied for the synthesis of $Fe_2O_3$ nanowires (Tang et al. 2015). $Fe_2O_3$ nanowires achieved a high specific capacitance of 908 F $g^{-1}$ at 2 A $g^{-1}$ in a broad negative voltage of -1.35-0 V in 2M KOH aqueous solution. Yang et al (Yang et al. 2014) prepared amorphous $Fe_2O_3$ nanotubes on flexible carbon fabric employing ZnO nanorods as the template. The as-obtained amorphous $Fe_2O_3$ nanotubes delivered a large specific capacitance of 257.8 F $g^{-1}$ at 1.4 A $g^{-1}$ in 5 M LiCl electrolyte, which can be attributed to the novel nanotubular structure with large interfacial areas and shortened pathways for ion diffusion.

Compared with 0D and 1D nanostructures, 2D nanostructures such as nanosheets and nanoflakes are more helpful in enlarging the contact area with electrolyte and shortening the pathways of ion diffusion, and hence usually receiving outstanding pseudocapacitive performance (Li et al. 2017; Liu et al. 2015). For instance, Liu et al. (Liu et al. 2015) prepared ultrathin $α$-$Fe_2O_3$ nanoflakes from $α$-$Fe_2O_3$ nanorods using the electrochemically induced method, as shown in **Figure 9g**. Compared with $α$-$Fe_2O_3$ nanorods-based electrode (NR), the obtained $α$-$Fe_2O_3$ nanoflakes-based electrode (NF) delivered a higher areal capacitance of 145.9 mF $cm^{-2}$ at 1 mA $cm^{-2}$ (**Figure 9h**), which is due to their increased surface area and decreased charge transfer resistance (**Figure 9i**).

In general, 3D porous nanostructures have some features such as large surface area and structural flexibility which can help to provide more contact area with electrolyte ion and efficiently accommodate volume change. Zheng et al. (Zheng, Yan, et al. 2016) fabricated $α$-$Fe_2O_3$ hollow nanoshuttles with a uniform wall thickness of 30 nm and a length of 100 nm using hydrothermal method (**Figure 9j**). **Figure 9k** display that the $α$-$Fe_2O_3$ hollow nanoshuttles achieved a high capacitance of 249 F $g^{-1}$ at 0.5 A $g^{-1}$. As shown in **Figure 9l**, there was only minor change in the specific capacitance of $α$-$Fe_2$O3 hollow nanoshuttles from 20 °C (203 F $g^{-1}$) to 60 °C (234 F $g^{-1}$).

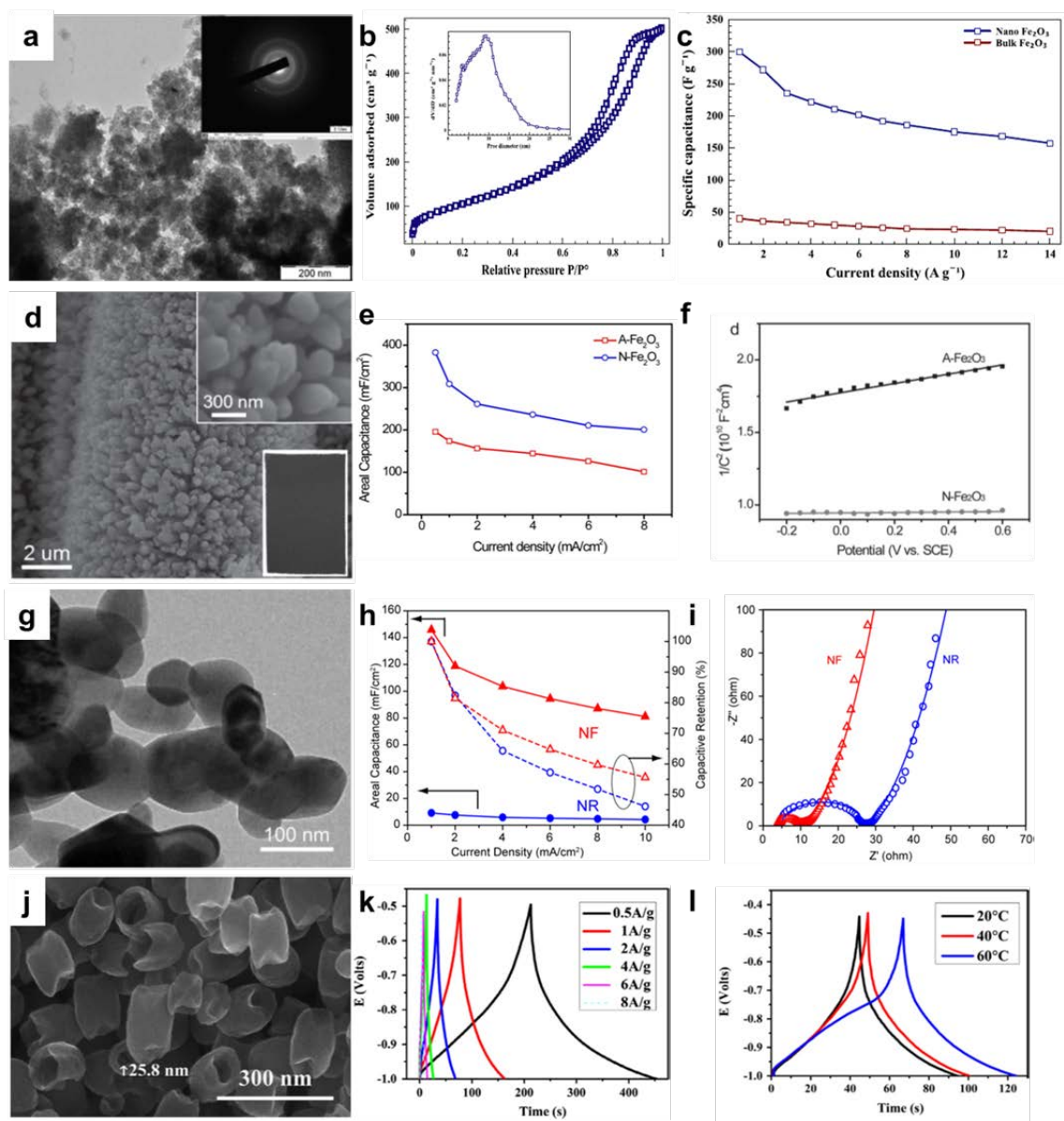

**Figure 8.** (a) TEM image and SAED pattern (inset), (b) N$_2$ adsorption-desorption isotherm and BJH pore-size distribution curve (inset), and (c) rate capability of α-Fe$_2$O$_3$ nanoparticles (Nano Fe$_2$O$_3$). Reproduced with permission (Shivakumara et al. 2014). Copyright 2014, Elsevier. (d) FE-SEM image, (e) rate capability, and (f) Mott-Schottky plots of oxygen-deficient α-Fe$_2$O$_3$ nanorods (N-Fe$_2$O$_3$). Reproduced with permission (Lu et al. 2014). Copyright 2014, Wiley. (g) TEM image, (h) rate capability, and (i) Nyquist plots of α-Fe$_2$O$_3$ nanoflakes (NF). Reproduced with permission (Liu et al. 2015). Copyright 2015, Elsevier. (j) FE-SEM image, (k) GCD curves at different current densities, (l) GCD curves at different temperatures of α-Fe$_2$O$_3$ hollow nanoshuttles. Reproduced with permission (Zheng, Yan, et al. 2016). Copyright 2016, Elsevier.

Fe$_3$O$_4$ with various morphologies were also prepared and studied for SCs. It is worth noting that the molar ratio of Fe$^{2+}$/Fe$^{3+}$ should be 0.5 for the synthesis of Fe$_3$O$_4$ nanoparticles (Mo et al. 2011; Guan et al. 2012). However, the obtained Fe$_3$O$_4$ nanoparticles are easy to aggregate, leading to large particles and small specific surface area. Wang et al. (Wang et al. 2013) prepared $Fe_3O_4$ nanoparticles with 2 μm size using the ultrasonic-assisted method in ethanolamine solvent. The as-prepared $Fe_3O_4$ nanoparticles possessed a large specific surface area of 165 $m^2$ $g^{-1}$ due to the introduction of ethanolamine and ultrasound, leading to a high specific capacitance of 207.7 F $g^{-1}$ at 0.4 A $g^{-1}$. Recently, $Fe_3O_4$ with 2D structures were successfully prepared (Li, Chen, et al. 2016). $Fe_3O_4$ nanosheet arrays grown on Ni foam delivered a high specific capacitance of 379.8 F $g^{-1}$ at 2 A $g^{-1}$ in 2 M KOH electrolyte, which can be attributed to their 2D porous structure and good electrical conductivity.

**3.2 Intercalation dominated pseudocapacitive electrode materials**

**3.2.1** *Vanadium oxides*

As typical pseudocapacitive electrode materials, vanadium oxides ($V_2O_5$, $VO_2$, $V_2O_3$) have been widely studied (Liu, Su, et al. 2017; Wu et al. 2013; Yan, Li, et al. 2016). However, the electrochemical properties of vanadium oxides are limited due to their poor electrical conductivity ($10^{-2}$-$10^{-3}$ S $cm^{-1}$) and structural instability (Wang et al. 2018; Yan, Li, et al. 2016). Therefore, in order to improve the electrochemical performance of vanadium oxides-based electrode materials, many methods have been proposed such as hybridization with highly conductive materials and structural control. Vanadium oxides-based composites are presented in the next chapter in detail. In this section, pure vanadium oxides with different morphologies and pseudocapacitive properties are introduced.

For orthorhombic $V_2O_5$, it has a layered structure with square pyramid-like $VO_6$ octahedral units, which can be used as intercalation pseudocapacitive material, as shown in (**Figure 10a**) (Rui et al. 2013). The electrochemical reaction mechanism of $V_2O_5$ in neutral aqueous electrolytes can be expressed as **equation (12)** (Saravanakumar et al. 2012):

$V_2O_5 + xM^+ + xe^- \leftrightarrow M_xV_2O_5$ 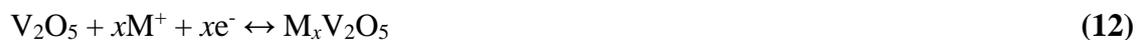 **(12)**

where M represents Li, Na, or K. In order to accelerate charge transfer and electrolyte ion diffusion, different synthesis methods have been employed to control the $V_2O_5$ nanostructures. $V_2O_5$ shows strong preferential growth orientation along the long axis. Therefore, the preparation of 0D $V_2O_5$ architectures needs appropriate strategic and parametric design (Liu, Su, et al. 2017). Thermal pyrolysis is a highly effective approach for the synthesis of $V_2O_5$. 0D porous $V_2O_5$ nanoparticles were prepared by the conversion of $(NH_4)_2V_3O_8$/carbon composites (**Figure 10d**) (Zhang, Zheng, et al. 2017). Compared aqueous electrolyte, $V_2O_5$ nanoparticles exhibited a higher specific capacitance of 545 F·g$^{-1}$ at a current density of 1 A·g$^{-1}$ in the organic electrolytes. 1D nanostructures have gained extensive attention due to their unique structure features such as large surface-to-volume ratio and easy to be linked as networks. Hydrothermal method is an effective technique to prepare $V_2O_5$ nanowires. For example, Qian et al. (Qian, Zhuo, et al. 2015) designed $V_2O_5 \cdot H_2O$ 1D nanowires using the hydrothermal method and PEG-6000 as a soft template (**Figure 10e**). The pseudocapacitive behaviors of $V_2O_5 \cdot H_2O$ 1D nanowires displayed outstanding pseudocapacitance of 349 F g$^{-1}$ at 5 mV s$^{-1}$ but poor cycling stability (capacitance retention of 20% after 200 cycles), as shown in **Figure 10f,g**. In recent years, 2D nanostructures have attracted considerable attention due to their unusual properties. Nagaraju et al. (Nagaraju et al. 2014) have reported 2D $V_2O_5$ nanosheets using hydrothermolysis method for the application of SCs (**Figure 10h,i**). The specific capacitance of 2D $V_2O_5$ nanosheets can reach 253 F g$^{-1}$ at a current density of 1 A g$^{-1}$ in 1 M KCl electrolyte (**Figure 10j**). For hierarchical nanostructures, the assembling architectures have significant impacts on their pseudocapacitive performance. 3D $V_2O_5$ structures assembled by nanosheets with a thickness of 4 nm were designed, which achieved a large surface area of 133 m$^2$ g$^{-1}$ and an ultrahigh specific capacitance of 521 F g$^{-1}$ at 5 mV s$^{-1}$ (**Figure 9k-m**) (Zhu et al. 2013).

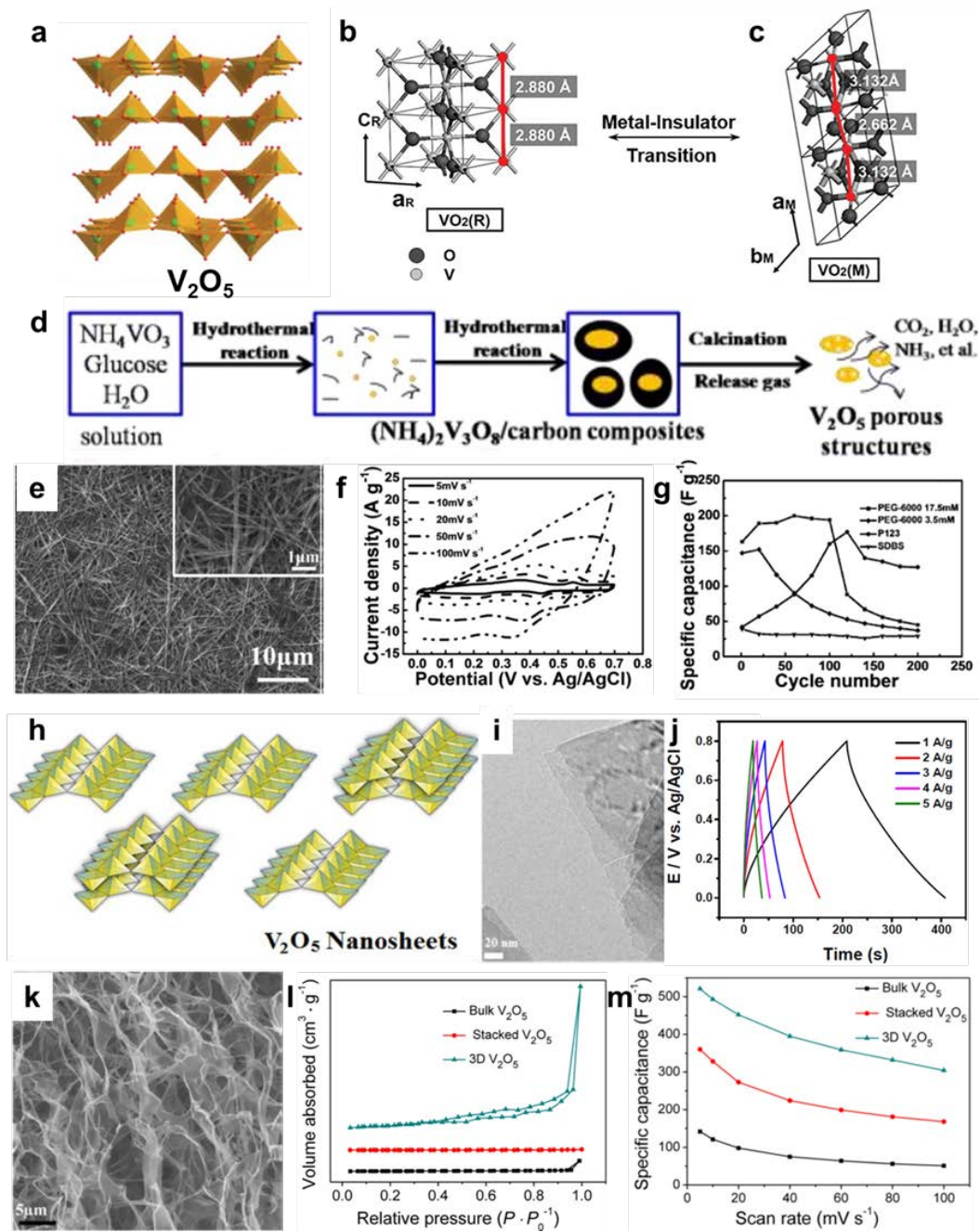

**Figure 9.** (a) Crystal structure of bulk $V_2O_5$. Reproduced with permission (Rui et al. 2013). Copyright 2013, Royal of Society of Chemistry. (b) Crystal structures of $VO_2(R)$ and (c) $VO_2(M)$. Reproduced with permission (Li, Ji, et al. 2013). Copyright 2013, Springer Nature Publishing. (d) Schematic diagram of the synthesis process of $V_2O_5$ porous nanoparticles. Reproduced with permission (Zhang, Zheng, et al. 2017). Copyright 2017, Elsevier. (e) SEM image, (f) CV curves at different scan rates, and (g) cycling stability of $V_2O_5$ nanowires (PRG-6000 17.5mM). Reproduced with permission (Qian, Zhuo, et al. 2015). Copyright 2015, Wiley. (h) Crystal structure, (i) TEM image, and (j) GCD curves at different current densities of $V_2O_5$ nanosheets. Reproduced with permission (Nagaraju et al. 2014). Copyright 2014, Royal of Society of Chemistry. (k) SEM image, (l) $N_2$ adsorption-desorption isotherm curve, and (m) specific capacitance at different scan rates of 3D $V_2O_5$ nanostructure. Reproduced with permission (Zhu et al. 2013). Copyright 2013, American Chemical Society.

Among various vanadium oxide-based materials, $VO_2$ is also favored by some researchers due to their high electronic conductivity and structural stability (Yan, Li, et al. 2016). For $VO_2$, there are several crystal structures, including monoclinic $VO_2(M)$, rutile $VO_2(R)$, mineral phases such as paramontroseite and nsutite-type $VO_2$, and some metastable phases such as $VO_2(A)$, $VO_2(B)$, $VO_2(C)$, and $VO_2(D)$ (Liu, Su, et al. 2017). Among various polymorphs, $VO_2(M)$ and $VO_2(R)$ can transform each other, as shown in **Figure 10b,c** (Li, Ji, et al. 2013). $VO_2$ with different architectures have been prepared for SCs (Pan et al. 2013; Xia et al. 2015; Shao et al. 2012). For example, 0D $VO_2$ nanoparticles with a size of around 300 nm were obtained by the assistance of $H_2$ treatment (Pan et al. 2013). After $H_2$ thermal treatment, the conductivity of $VO_2$ was increased nearly three times because of the synergies of H-doping, oxygen vacancies, and slight reduction of V, which was consistent with the ESR results (0.8 Ω for $H_2$ treated $VO_2$, 2.5 Ω for untreated $VO_2$). Compared with untreated $VO_2$ (76 F $g^{-1}$), $H_2$ treated $VO_2$ nanoparticles can achieve a higher specific capacitive of 300 F $g^{-1}$ at 1 A $g^{-1}$. 2D nanostructures can be directly grown on the surface of the flexible conductive collector. Xia et al. (Xia et al. 2015) designed an integrated electrode composed of graphene foam (GF) and $VO_2$ nanoflakes. The binder-free electrode $VO_2$/GF exhibited a high rate capability (350 F $g^{-1}$ at 2 A $g^{-1}$ and 170 F $g^{-1}$ at 32 A $g^{-1}$), which can be attributed to the integrated porous conductive architecture. Shao et al. (Shao et al. 2012) designed 3D hexangular starfruit-like $VO_2$ by a one-step hydrothermal method in the presence of P123 (poly(ethylene oxide)-block-poly(propylene oxide)-block-poly(ethylene oxide)). P123 plays a crucial role in the assembly of 3D nanostructures. 3D hexangular starfruit-like $VO_2$ obtained a specific capacitance of 218 F $g^{-1}$ at a scan rate of 5 mV $s^{-1}$.

For $V_2O_3$, it has metallic behavior due to the itinerating of V 3d electrons along the V-V chains. Besides the above feature, $V_2O_3$ with unique tunneled structure can facilitate the intercalation/deintercalation of ions (Yan, Li, et al. 2016). However, few kinds of literature about $V_2O_3$ for the SC applications are reported, which is because their synthesis conditions are harsh. Liu et al. prepared 3D hierarchical flower-like $V_2O_3$ nanostructure composed of 2D nanoflakes by solvothermal-heating treatment method

(Liu et al. 2010). The 3D flower-like V$_2$O$_3$ displayed a specific capacitance of 218 F g$^{-1}$ at a current density of 0.05 A g$^{-1}$ in Li$_2$SO$_4$ electrolyte.

**3.2.2** *Tungsten oxides*

Recently, tungsten oxide (WO$_3$) is regarded as promising candidates for SCs due to its good electronic conductivity (10$^{-6}$-10 S cm$^{-1}$), environmental friendliness, low cost, and high theoretical specific capacitance of 1112 F g$^{-1}$ (Zheng et al. 2017; Liu, Sheng, et al. 2018; Yang, Sun, et al. 2015; Zhu et al. 2014). Besides the above features, WO$_3$ has many polymorphs including tetragonal *α*-WO$_3$, orthorhombic *β*-WO$_3$, monoclinic *γ*-WO$_3$, triclinic *δ*-WO$_3$, monoclinic *ε*-WO$_3$, and hexagonal *h*-WO$_3$ (Cong et al. 2016). Among these polymorphs of WO$_3$, *h*-WO$_3$ possesses hexagonal tunnels and tetragonal tunnels, which enable the insertion of protons without changing the crystal structure, leading to a intercalation pseudocapacitive behavior (WO$_3$ + $x$H$^+$ + $x$e$^-$ ↔ H$_x$WO$_3$, 0 < $x$ < 1), as shown in **Figure 11a** (Zhu et al. 2014). *h*-WO$_3$ with 3D pillar-like nanostructure can achieve a high capacitance of 421.8 F g$^{-1}$ at a current density of 0.5 A g$^{-1}$ (**Figure 11b,c**) (Zhu et al. 2014). Jia et al. (Jia et al. 2018) prepared *h*-WO$_3$ with 2D mesoporous pancake-like structure using the hydrothermal method without capping agent (**Figure 11d**). *h*-WO$_3$ nanopancakes obtained a specific capacitance of 605.5 F g$^{-1}$ at 0.37 A g$^{-1}$ and high cycling stability (capacitance retention of 110.2% after 4000 cycles). For WO$_3$ with other crystal structures, they must be distributed in/on the carbon-based materials, otherwise, the WO$_3$ will suffer from the redox phase transformations and display battery behavior (Liu, Sheng, et al. 2018; Wang, Wang, et al. 2014; Liu, Sheng, et al. 2018). The intercalation dominated pseudocapacitive behavior of WO$_3$ can be maintained via providing proton transport pathways by the assistance of structural water. As shown in **Figure 11e**, benefiting from the structural water, the hydrated WO$_3$ (WO$_3$·2H$_2$O) has a layered structure, which can accelerate the intercalation/deintercalation of protons (Mitchell et al. 2017). However, there is no layered structure in the anhydrous monoclinic WO$_3$ (*γ*-WO$_3$) without structural water (**Figure 11e**), leading to a battery-type behavior. CV curves of WO$_3$·2H$_2$O clearly demonstrate its intercalation pseudocapacitive behavior, and CV curves of *γ*-WO$_3$ display its battery behavior (**Figure 11f,g**).

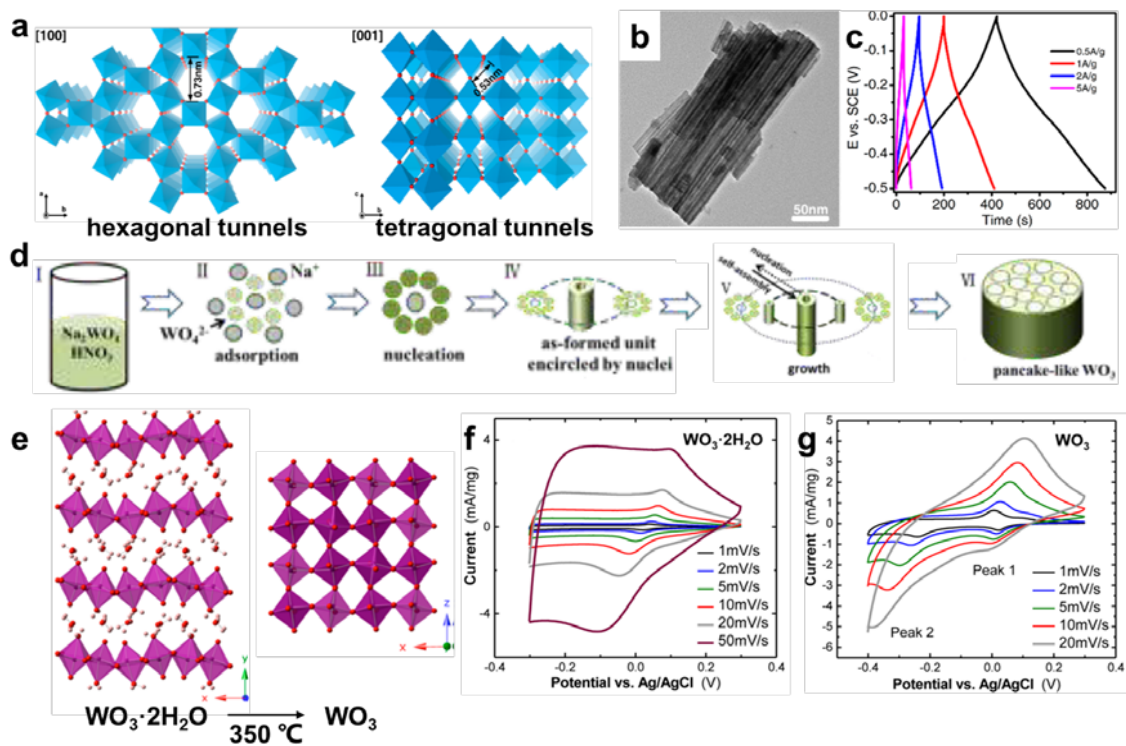

**Figure 11.** (a) Schematic diagram of the *h*-WO$_3$ along the [100] and [001] directions with hexagonal tunnels and tetragonal tunnels. (b) TEM image 3D *h*-WO$_3$ nanopillars and (c) GCD curves of 3D *h*-WO$_3$ nanopillars at various current densities. Reproduced with permission (Zhu et al. 2014). Copyright 2014, American Chemical Society. (d) Schematic diagram of the synthesis process of 2D *h*-WO$_3$ nanopancakes. Reproduced with permission (Jia et al. 2018) Copyright 2018, Royal of Society of Chemistry. (e) Crystallographic structures of WO$_3$·2H$_2$O and *γ*-WO$_3$. CV curves of (f) WO$_3$·2H$_2$O and (g) *γ*-WO$_3$ at different scan rates. Reproduced with permission (Mitchell et al. 2017). Copyright 2017, American Chemical Society.

### 3.2.3 *Molybdenum oxides*

Recently, molybdenum oxides such as MoO$_3$ and MoO$_2$ have attracted significant interest as active materials for SCs due to their high specific capacitance, low cost, natural abundance, and environmental friendliness (Hu, Zhang, et al. 2015).

For MoO$_3$, it has two basic polymorphs including thermodynamically stable orthorhombic *α*-MoO$_3$ and metastable monoclinic *β*-MoO$_3$ with a ReO$_3$-type structure (Hu, Zhang, et al. 2015). *α*-MoO$_3$ is a typical 2D layered material with an electrochemical activity that can accommodate up to 1.5 Li/Mo, and comprises alternately stacked layers held together by weak van der Waals forces along [010], which enable the insertion of electrolyte cations without changing the crystal structure, as shown in **Figure 12a** (Hashem et al. 2012). Therefore, $\alpha$-MoO$_3$ is able to be an intercalation pseudocapacitive material. In order to demonstrate the pseudocapacitive behavior in $\alpha$-MoO$_3$ with a layered structure, Brezesinski et al. (Brezesinski et al. 2010) prepared ordered mesoporous $\alpha$-MoO$_3$ thin film by an evaporation-induced self-assembly method (**Figure 12b**). It was found that the specific capacitance of mesoporous $\alpha$-MoO$_3$ was higher than those of either mesoporous amorphous material or non-porous crystalline $\alpha$-MoO$_3$ (**Figure 12c**). Both mesoporous amorphous and non-porous crystalline MoO$_3$ showed redox pseudocapacitance, but mesoporous $\alpha$-MoO$_3$ enabled Li$^+$ to be inserted into the van der Waals gaps of $\alpha$-MoO$_3$, leading to the introduction of a unique intercalation pseudocapacitance. $\alpha$-MoO$_3$ with other morphologies including 1D nanorods and nanowires (Shakir et al. 2010; Liang et al. 2011), and 2D nanoplates and nanobelts (Tang et al. 2011; Jiang et al. 2013), have been reported as electrode materials for SCs, and exhibited improved pseudocapacitive properties.

For MoO$_2$, its crystal structure is shown in **Figure 12d** (Hu, Zhang, et al. 2015). MoO$_2$ octahedra composed of monoclinic cells are distorted and Mo atoms are off-center, leading to alternating short and long Mo-Mo distance and Mo-Mo bonding (Mackay and Henderson 2002). MoO$_2$ also possesses low metallic electrical resistivity ($8.8 \times 10$ $\Omega$ cm at 300 K in bulk MoO$_2$) and outstanding chemical stability, which enable it to be a promising candidate for SCs (Rogers et al. 1969). Various MoO$_2$ nanostructures have been designed for SC application (Zhou et al. 2016; Rajeswari et al. 2009; Xuan et al. 2016; Li, Shao, et al. 2013). For example, Zhou et al. (Zhou et al. 2016) synthesized 0D MoO$_2$ nanoparticles with the size of 200 nm by hydrothermal method. The MoO$_2$ nanoparticles showed a high specific capacitance of 621 F g$^{-1}$ at 1 A g$^{-1}$ and excellent cyclic performance (capacitance retention of 90% after 1000 cycles). MoO$_2$ with 1D rod-like structure was fabricated by thermal decomposition of tetrabutylammonium hexamolybdate (((C$_4$H$_9$)$_4$N)$_2$Mo$_6$O$_{19}$) in the atmosphere (Rajeswari et al. 2009). 1D MoO$_2$ nanorods showed a good specific capacitance of 140 F g$^{-1}$ at 1mA cm$^{-2}$ in 1 M H$_2$SO$_4$ solution. 2D MoO$_2$ nanoplates

were successfully synthesized by thermal decomposition of ammonium paramolybdate (($NH_4$)$_6$$Mo_7O_{24}$·$4H_2O$) (Xuan et al. 2016). $MoO_2$ nanoplates achieved a specific capacitance of 318 F g$^{-1}$ at 0.5A g$^{-1}$. Li et al. (Li, Shao, et al. 2013) prepared ordered mesoporous $MoO_2$ using silica KIT-6 as hard template (**Figure 12e**). By electrochemical quartz crystal microbalance test (EQCM), the dominating working ions for the charge storage of mesoporous $MoO_2$ in aqueous LiOH electrolyte were proved to be Li$^+$, as shown in **Figure 12f**. The mesoporous $MoO_2$ exhibited good cycling performance (capacitance retention of 90% after 1000 cycles) due to their unique mesoporous structure and low solubility in alkaline LiOH solution (**Figure 12g,h**).

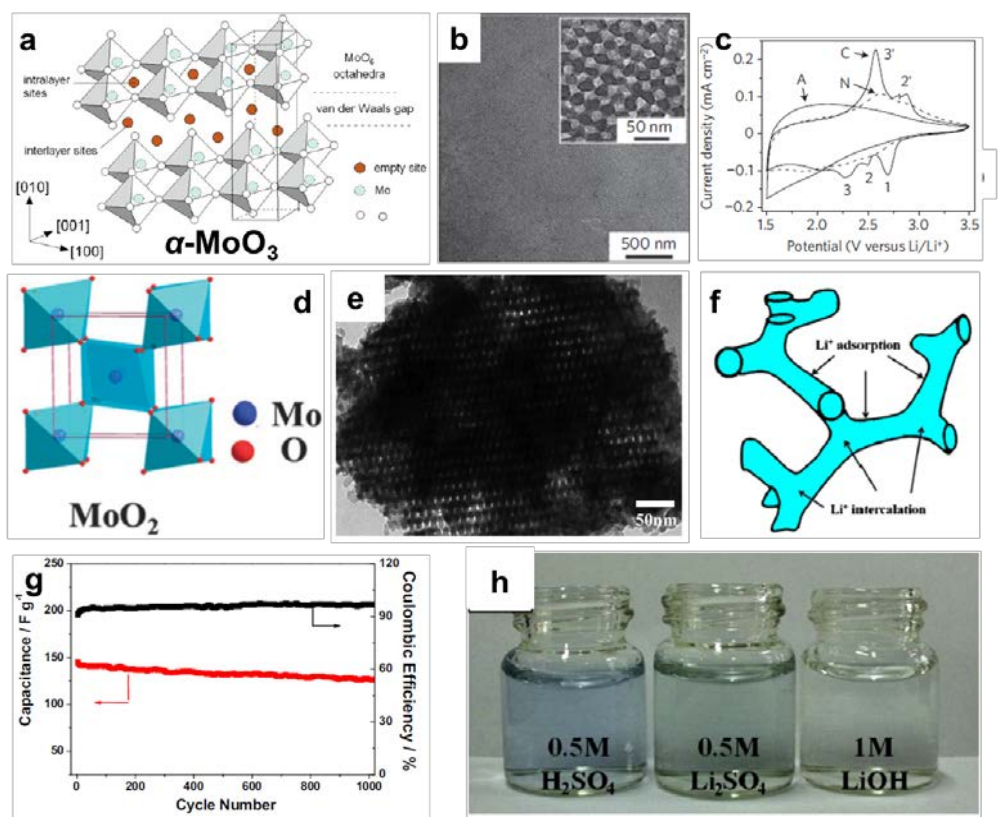

**Figure 11.** (a) Cyrstal structure of α-$MoO_3$. Reproduced with permission (Hashem et al. 2012). Copyright 2012, Elsevier. (b) Low-magnification bright-field TEM micrograph of mesoporous α-$MoO_3$ (inset: a higher magnification micrograph) and (c) cyclic voltammograms for different $MoO_3$ systems at a sweep rate of 1mV s$^{-1}$ (A: mesoporous amorphous $MoO_3$, C: the mesoporous crystalline α-$MoO_3$, and N: non-porous crystalline $MoO_3$). Reproduced with permission (Brezesinski et al. 2010). Copyright 2010, Springer Nature Publishing. (d) Cyrstal structure of monoclinic $MoO_2$. Reproduced with permission (Hu, Zhang, et al. 2015). Copyright 2015, Royal of Society of Chemistry. (e) TEM image of mesoporous $MoO_2$, (f) schematic illustration of the charge storage of mesoporous $MoO_2$ including intercalation and adsorption of Li$^+$, (g) cycling performance of mesoporous $MoO_2$ at the current density of 1 A g$^{-1}$, and (h) dissolution results of mesoporous $MoO_2$ in three types of electrolyte. Reproduced with permission (Li, Shao, et al. 2013). Copyright 2013,

Elsevier.

**3.2.4** *Niobium oxides*

Niobium oxide (Nb$_2$O$_5$) as intercalation pseudocapacitive electrode material has attracted great interests due to its outstanding structural stability and high theoretical capacitance (Yan, Rui, et al. 2016). The charge storage mechanism of Nb$_2$O$_5$ in Li-ion SCs can be described as **equation (13)**:

$$Nb_2O_5 + xLi^+ + xe^- \leftrightarrow Li_xNb_2O_5 \ (x \leq 2) \tag{13}$$

Nb$_2$O$_5$ has many thermodynamic stable polymorphic forms including pseudohexagonal (TT-Nb$_2$O$_5$), orthorhombic (T-Nb$_2$O$_5$), and monoclinic (H-Nb$_2$O$_5$) (Zhao et al. 2012). Among them, T-Nb$_2$O$_5$ is a promising candidate for the intercalation of Li ions because of its layered crystal structure as shown in **Figure 13a** (Augustyn et al. 2013). The tunnels of T-Nb$_2$O$_5$ offered by empty octahedral sites between (001) planes can serve as pathways for the intercalation of Li$^+$. Kim et al. (Kim et al. 2012) prepared 0D T-Nb$_2$O$_5$ nanoparticles using an aqueous sol-gel technique (**Figure 12b**). Compared with amorphous Nb$_2$O$_5$ (262 F g$^{-1}$, 4 min), T-Nb$_2$O$_5$ (555 F g$^{-1}$, 4 min) received a higher specific capacitance though the amorphous Nb$_2$O$_5$ has a much higher specific surface area (**Figure 12c**). The intercalation pseudocapacitors do not need high surface areas to obtain high pseudocapacitance, because the contact surfaces of T-Nb$_2$O$_5$ and electrolyte do not strictly determine the high rate behavior. Therefore, such pseudocapacitive materials bring new opportunities to practical devices.

**3.2.5** *Titanium oxides*

Titanium oxide (TiO$_2$) has many polymorphs including anatase, rutile, brookite, and TiO$_2$(B) (bronze). Among them, TiO$_2$(B) has attracted great attention as an electrode material for intercalation dominated pseudocapacitors, owing to its lower density and more open structure (Zukalova et al. 2005). As shown in **Figure 13d**, TiO$_2$(B) is composed of two edge-sharing TiO$_6$ octahedra linked at corners (Dylla et al. 2013). The charge storage mechanism in TiO$_2$(B) via the intercalation of Li$^+$ can be summarized as **equation (14)** (Zukalova et al. 2005):

$$TiO_2(B) + xLi^+ + xe^- \leftrightarrow 4Li_xTiO_2(B) \tag{14}$$

The main issues of TiO$_2$(B) are its intrinsic low conductivity and slow ion intercalation/deintercalation kinetics. Nanostructuring, in the architecture of 0D nanoparticles, 1D nanowires, and 2D nanosheets, can shorten the ion diffusion pathway and thus improve the rate capability even further (Dylla et al. 2013; Dylla et al. 2012). Dylla et al. (Dylla et al. 2012) prepared TiO$_2$(B) with different morphologies and compared their electrochemical properties. It was found that the specific capacity of TiO$_2$(B) nanosheets was higher than that of TiO$_2$(B) nanoparticles, which is attributed to the different lithium insertion behaviors (**Figure 13e**). DFT+U calculation results displayed that the different lithiation mechanisms are concerned with the elongated geometry of the nanosheet along the a-axis that shortens Li$^+$-Li$^+$ interactions between C and A2 sites (**Figure 13f**). TiO$_2$(B) is an example of intercalation pseudocapacitors. It displays broad redox peaks in CV curves and continuously sloping GCD curves, even in the bulk forms. However, compared with the bulk forms, the nanostructured TiO$_2$(B) present higher capacity and rate capability and exhibit irreversible capacity loss during the cycling process, which is an unresolved issue.

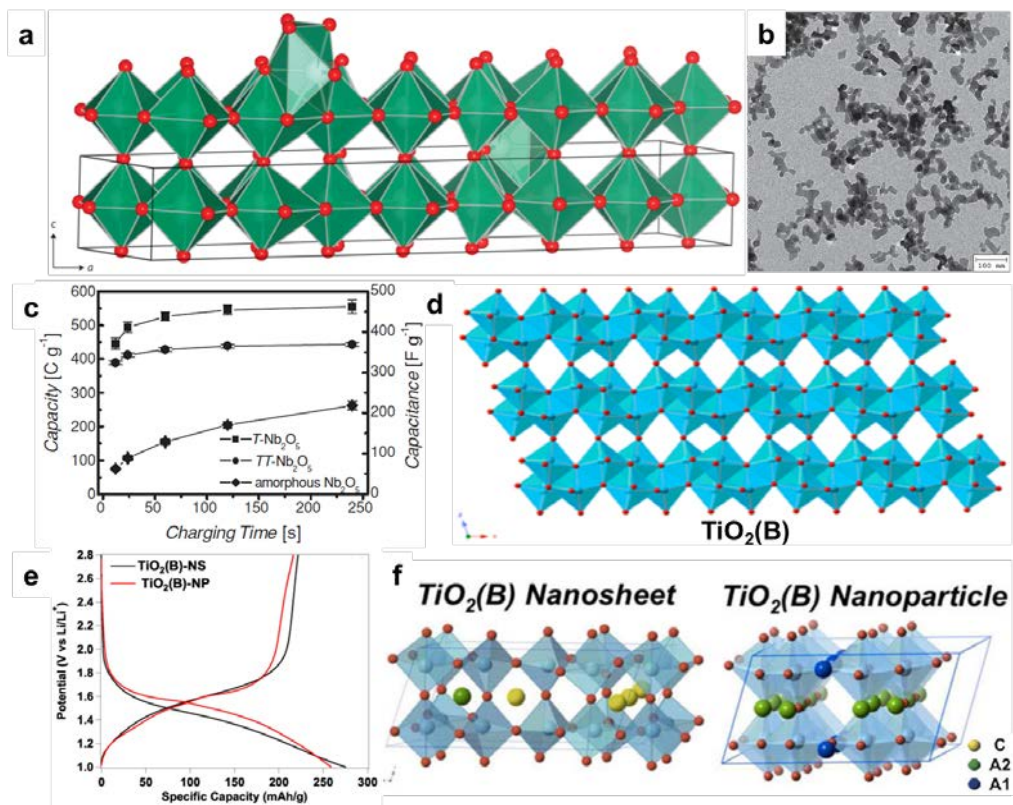

**Figure 13.** (a) Crystal structure of T-$Nb_2O_5$. Reproduced with permission (Augustyn et al. 2013). Copyright 2013, Springer Nature Publishing. (b) TEM image of T-$Nb_2O_5$ nanoparticles and (c) voltammetric sweeps of $Nb_2O_5$ in lithium-ion electrolyte. Reproduced with permission (Kim et al. 2012). Copyright 2012, Wiley. (d) Crystal structure of $TiO_2$(B). Reproduced with permission (Dylla et al. 2013). Copyright 2013, American Chemical Society. (e) Third cycle GCD curves of $TiO_2$(B) nanosheets (NS) and nanoparticles (NP) at 25 mA $g^{-1}$, (f) DFT+U calculated site occupancies in $TiO_2$(B) nanosheets and nanoparticles. Reproduced with permission (Dylla et al. 2012). Copyright 2012, American Chemical Society.

## 3.3 Battery-type materials

### 3.3.1 *Nickel oxides*

NiO has been attracted wide research attention and has been recognized as one of the promising candidates for SCs due to its high theoretical specific capacitance of 2584 F $g^{-1}$ (with a potential window of 0.5 V), resourcefulness, environment friendliness, and good thermal and chemical stability (Pang et al. 2012). The charge storage phenomenon including the redox reactions (main contribution) and the EDLC formed on the electrode/electrolyte interface. The performance of NiO pseudocapacitor electrode is determined by the redox reaction of nickel oxide in an alkaline electrolyte. The typical CV curve (Liang et al. 2012) for NiO is different from that of the EDLCs. Two obvious peaks indicate the faradic oxidation/reduction reactions. The oxidation peak at 0.46 V is assigned to the conversion of NiO to NiOOH, whereas the reduction peak at 0.045 V is identified as the reverse reaction as shown in **equation (15)**.

$$NiO + OH^- \leftrightarrow NiOOH + e^- \tag{15}$$

Although NiO possesses extremely high theoretical capacitance, this value has not been achieved experimentally due to its poor electrical conductivity (0.01-0.32 S $m^{-1}$). Until now, the typical practical capacitance based on the NiO based electrodes is within the range of 50-1776 F $g^{-1}$ (Lu, Lattanzi, et al. 2011; Wang et al. 2010; Liang et al. 2012). The structure is found to largely influence the SC of electrode materials, the parameters of which include morphology, size, specific surface area, pore size, and distribution. The morphology largely determines the specific surface area. Some systematic reports on the relationship between the unique morphology-dependent supercapacitance properties and the surface characteristics gave some results not the

same as we expect. The measured SC is not changed linearly with the change of the specific surface area. This phenomenon probably because of that some of the surface areas is not electrochemically accessible by the electrolyte. The porosity itself is not a simple parameter, relating to both pore sizes and pore-size distribution for a given overall specific area of the material. So the porosity is one of the key factors for the development of high capacity. Based on Lu's research, the performance of supercapacitors will be influenced by the size, geometry, length, alignment of pores and the degree of pore ordering (Lu, Chen Xiao 2013). Therefore, the capacitance largely depends on the exposed surface of the electrode accessible to the electrolyte.

The ion diffusion rate and electrolyte wettability are closely related to materials the morphology and particle size of electrode materials. Micro/nanostructure materials are wildly studied to be applied for the electrochemical capacitors because of fast redox reactions, shortened diffusion paths and high specific surface areas in the solid phase. NiO nanostructures including 1D (e.g. nanotubes or nanowires), 2D (nanoplatelets or nanosheets) and 3D cross-linked materials have been fabricated and their electrochemical performance has been studied via various electrochemical examinations.

1D nanostructures are perfect building blocks for functional nanodevices and represent the smallest dimension for efficient electron and exciton transport (Huang et al. 2001). Hence 1D nanostructures are expected to reduced contact resistance and facilitate the charge/discharge kinetics and show better charge transportability. Mesoporous NiO nanowires synthesized by a hydrothermal reaction and subsequent annealing at 400 °C exhibited a high specific capacitance of 348 F $g^{-1}$ as electrode materials for electrochemical capacitors (Su et al. 2012). NiO nanorod arrays on Ni foam prepared by Lu et al. have been found to show ultrahigh specific capacitance (2018 F $g^{-1}$ at 2.27 A $g^{-1}$), high power density (1536 F $g^{-1}$ at 22.7 A $g^{-1}$), and good cycling stability (only 8% of capacitance was lost in the first 100 cycles with no further change in the subsequent 400 cycles) (Lu, Chang, et al. 2011). This work improved the reversible capacitance record of NiO by 50% or more, achieving 80% of its theoretical value, indicating that 1D ordered porous array on 3D substrate structure can provide

multiple effects for high-performance capacitors. The excellent performance was due to the thin (< 20 nm) rod structure, high crystallinity, regularly aligned array arrangement and strong bonding of the nanorods to the metallic Ni substrate. 1D nanotube structure shows high porosity, increased surface area, and low mass density. Cao and partner (Cao et al. 2014) fabricated 3D hierarchical porous NiO nanotube arrays on nickel foam through a successive electro-deposition method combined with ZnO nanorod template. The diameter of the as-prepared porous NiO nanotubes is ~170 nm with ~10 nm thickness of interconnected branch nanoflakes. Benefited by the unique architecture, the NiO nanotube arrays demonstrate a high capacitance of 675 F $g^{-1}$ at the 2 A $g^{-1}$ and 569 F $g^{-1}$ at 40 A $g^{-1}$, respectively, as well as good cycling stability.

Wu et al. (Wu, Hui, et al. 2016) fabricated two-dimensional NiO nanoflake arrays vertically grown on the surface of three-dimensional nickel foam via a solvothermal reaction followed by sintering in air (**Figure 14a-c**). The nanoflakes are found to have an ultrathin thickness as thin as ~7 nm and abundant nanoscale pores (< 10 nm). The unique porous structure decreases diffusion resistance of electrolytes in fast redox reactions and preserves mechanical integrity during prolong charging/discharging process. The 2D ultrathin porous NiO nanoflakes electrode shows remarkably high specific capacitance (2013.7 F $g^{-1}$ at 1 A $g^{-1}$ and 1465.6 F $g^{-1}$ at 20 A $g^{-1}$) and excellent cycling ability (100% capacitance retention over 5000 cycles). The ultrathin porous NiO nanoflakes as a positive electrode are assembled with reduced graphene oxide (rGO) as a negative electrode to form an asymmetric supercapacitor (ASC) cell. The NiO//rGO ASC, operating at 1.5 V, displays a high specific capacitance of 145 F $g^{-1}$ at 1 A $g^{-1}$ with a high energy density of 45.3 Wh $kg^{-1}$ at a power density of 1081.9 W $kg^{-1}$ and excellent cycling stability (retain 91.1% of capacitance after 5000 cycles).

Porous hollow nanospheres assembled by NiO nanosheets have been synthesized by Ding and Li's group (**Figure 14d-f**) (Yu, Jiang, et al. 2014). For synthesis, polystyrene nanospheres with carboxyl groups (CPS) were initially prepared. Then after a two-step activation process, Ni was deposited on the CPS core accompanying with nucleation and growth of nanocrystals. CPS@Ni core-shell nanoparticles were formed in this step. After calcination at high temperature, the CPS core was removed and metallic Ni

nanoshell is oxidized to NiO in the air condition. The as-synthesized hollow NiO nanospheres a high reversible capacitance of 600 F g$^{-1}$ after 1000 cycles at a high current density of 10 A g$^{-1}$. The randomly oriented NiO nanosheets in the hollow spheres provided porous network structures for enhanced OH$^-$ transport, resulting in outstanding electrochemical performance.

Cheng et al. (Cheng et al. 2013) used carbon paper as the scaffolds for supporting the formation of a hybrid-structured NiO composed of nanonet and nanoflower structures (**Figure 14g-i**). The hybrid nanonet/nanoflower NiO was help for the fast ion and electron transfer, which was able to make full use of electrochemical charge storage of NiO electrode. High areal capacitance of 0.93 F cm$^{-2}$ (~840 F g$^{-1}$) was obtained due to the rapid charge transfer and ion diffusion, as well as high rate capability, and good cycling stability (keeping 70% of the initial capacitance after 10000 cycles).

Although morphology and particle size of NiO is related to surface area and the structure of nanomaterials for ions or electron transporting paths, the effect of these factors on the SC value only work within a certain range. The SC value is the result of a combination of aspects of morphology, particle size, surface area and pore properties of the material.

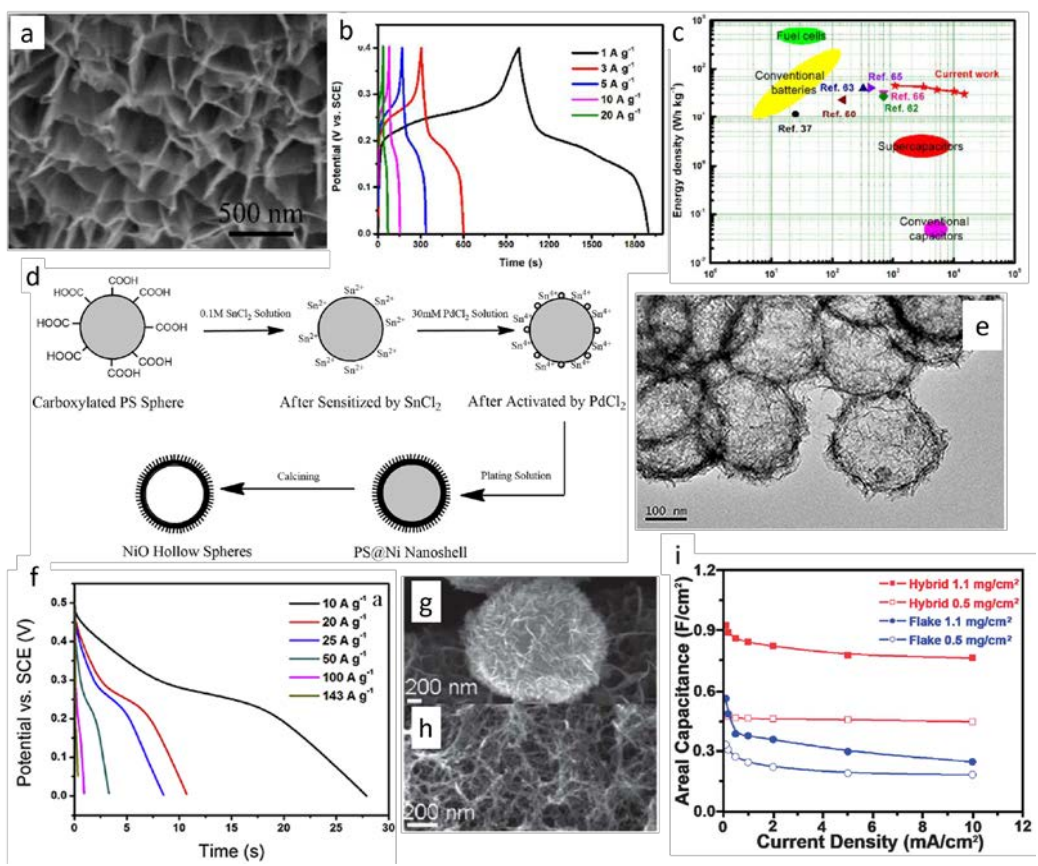

**Figure 14.** (a) SEM image of ultrathin porous NiO nanoflakes on nickel foam. (b) GCD curves of the ultrathin porous NiO nanoflakes electrode at various current densities. (c) Ragone plot of NiO//rGO compared with other reported data. Reproduced with permission (Wu, Hui, et al. 2016). Copyright 2016, Royal of Society of Chemistry. (d) Scheme for electroless deposition of Ni onto a CPS sphere and formation of a NiO nanosheet hollow sphere. (e) TEM image of NiO hollow sphere. (f) Galvanostatic discharge curves of NiO hollow spheres at various discharge current densities. Reproduced with permission (Yu, Jiang, et al. 2014). Copyright 2014, Elsevier. (g) High-magnification nano-flower. (h) High-magnification nanowire network (nanonet) on a carbon fiber. (i) Areal capacitances of CFP-supported NiO hybrid and flake electrodes with mass loadings of 0.5 and 1.1 mg cm$^{-2}$ at different current densities. Reproduced with permission (Cheng et al. 2013). Copyright 2013, Royal of Society of Chemistry.

### 3.3.2 *Cobalt oxides*

$Co_3O_4$ is the $AB_2O_4$ spinel structure belonging to a cubic system with a lattice constant $a = 0.808$ nm (JCPDS 42-1467). $Co_3O_4$ is considered as a highly promising material for supercapacitors due to its low cost, non-toxic, easy synthesis, environmentally friendly nature, and more importantly high theoretical capacitance (3560 F g$^{-1}$) (Liu et al. 1999). The pseudocapacitance of $Co_3O_4$ originates from the following redox reaction (Shan and Gao 2007; Xia et al. 2012):

$$Co_3O_4 + OH^- + H_2O \leftrightarrow 3CoOOH + e^- \quad (16)$$

$$CoOOH + OH^- \leftrightarrow CoO_2 + H_2O + e^- \quad (17)$$

So far, the synthesis of cobalt oxides has been developed various strategies including the sol-gel process, physical vapor deposition, chemical precipitation, hydro/solvo-thermal method, and electrochemical deposition (Liu, Long, et al. 2013). Some characteristics such as morphology, structures, and dimension can be simply adjusted by controlling the synthetic parameters, e.g. reaction temperature, reaction time, the concentration of matrix solution, a complexing agent, etc. (Wei et al. 2015). Because of the low electrical conductivity and small surface area, the observed specific capacitances of practical bulk $Co_3O_4$ pseudo-capacitors are far less than the theoretical prediction, especially at high current densities. Fabrication of a variety of morphologies and microstructures for $Co_3O_4$ by electrochemical and/or chemical approaches, such as nanoparticles (Yuan et al. 2011), nanowires (Xia et al. 2011), nanorods (Zheng, Li, et al. 2016), nanosheets (Kung et al. 2012), and porous nanostructures (Xiong et al. 2009), behave a profound influence on the supercapacitive performance owing to dissimilarities in the electrode/electrolyte interface features and ion transport rates during the charge storage processes. For example, Yadav et al. (Yadav and Chavan 2017) synthesized porous $Co_3O_4$ thin films with cubic phase by chemical spray pyrolysis using mixed aqueous/organic solvent. The studies on the substrate temperature revealed that $Co_3O_4$ deposited at 350 °C had a minimum electrical resistivity of $2.08 \times 10^3$ $\Omega$ cm. The maximum specific capacitance of $Co_3O_4$ thin films is up to 425 F $g^{-1}$ at a scan rate of 5 mV $s^{-1}$. The noteworthy cycle stability of 92.56% has been observed at a current density of 1 A $g^{-1}$ after 1000 charge/discharge cycles. To increase the electronic conductivity in advance, vertically aligned graphene nanosheets (VAGNs) supported by carbon fabric was used as a superb backbone for loading the $Co_3O_4$ nanoparticles (**Figure 15a-c**) (Liao et al. 2015). The as-prepared composite showed an ultra-high specific capacitance of 3480 F $g^{-1}$, which was very close to the theoretical value of 3560 F $g^{-1}$. The all-solid-state symmetric supercapacitor device made with two pieces of the free-standing hybrid electrodes is flexible and suitable for different bending angles. The device delivered a high capacitance (580 F $g^{-1}$), good cycling ability (86.2% capacitance

retention after 20000 cycles), high energy density (80 Wh kg$^{-1}$), and high power density (20 kW kg$^{-1}$ at 27 Wh kg$^{-1}$).

1D Co$_3$O$_4$ nanowires can be formed by a simple hydrothermal method, which was in-situ grown on 3D graphene foam (**Figure 15d**) (Dong et al. 2012). The 3D graphene/Co$_3$O$_4$ composite was used as the monolithic free-standing electrode for supercapacitor application and delivered high specific capacitance of ~1100 F g$^{-1}$ at a current density of 10 A g$^{-1}$ with excellent cycling stability (**Figure 15e,f**). Besides this, Co$_3$O$_4$ nanowires also can be grown on carbon fiber paper collectors (Rakhi et al. 2012). The Co$_3$O$_4$ nanowires are self-organized into a brush-like morphology with the nanowires completely surrounding the carbon microfiber cores (**Figure 15g**), which enable each nanowire directly contact with the current collector. This advanced structure with both micro- and nanopores facilitates the infiltration of electrolyte, reduces cell resistance and fasten the reaction kinetics. In contrast, flower-like morphology will be obtained by using planar graphitized carbon paper as the collector, which can't ensure the contact between the nanowire and the collector. As a result, the SC of brush-like and flower-like materials are 1525 and 1199 F g$^{-1}$, respectively, at a current density of 1 A g$^{-1}$ (**Figure 15h**). In addition, brush-like Co$_3$O$_4$ delivered high specific power (71 kW kg$^{-1}$) and specific energy (81 Wh kg$^{-1}$) in two electrode configuration, while 55 Wh kg$^{-1}$ and 37 kW kg$^{-1}$ only obtained by flower-like Co$_3$O$_4$ (**Figure 15i**).

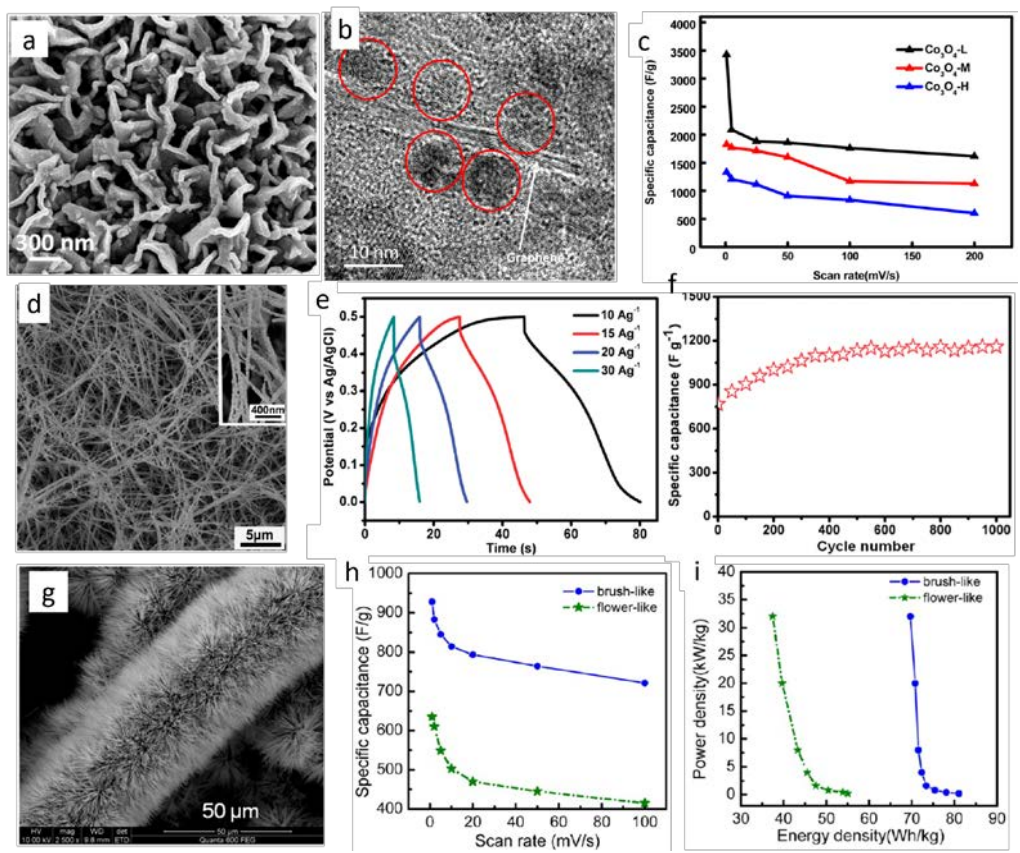

**Figure 15.** (a) SEM image of VAGN/Co$_3$O$_4$ hybrid. (b) High-resolution TEM image of the VAGN and Co$_3$O$_4$ particles. (c) Specific capacitance vs scan rate for various VAGN/Co$_3$O$_4$ hybrid with different Co$_3$O$_4$ loading amounts. Reproduced with permission (Liao et al. 2015). Copyright 2015, American Chemical Society. (d) SEM images of graphene/Co$_3$O$_4$ nanowire composite. Inset shows an enlarged view. (e) Galvanostatic charge/discharge curves of graphene/Co$_3$O$_4$ composite electrode at different current densities. (f) Cycling performance of graphene/Co$_3$O$_4$ composite electrode at a current density of 10.0 A g$^{-1}$. Reproduced with permission (Dong et al. 2012). Copyright 2012, American Chemical Society. (g) SEM image of Co$_3$O$_4$ nanowires with brush-like morphology. (h) Specific capacitances of Co$_3$O$_4$ nanowires with brush-like and flower-like morphologies at different scan rates. (i) Ragone plot (power density vs. energy density) of Co$_3$O$_4$ nanowires with brush-like and flower-like morphologies based symmetric supercapacitors. The energy densities and power densities were derived from the charge-discharge curves at different current densities. Reproduced with permission (Rakhi et al. 2012). Copyright 2012, American Chemical Society.

Kung et al. (Kung et al. 2012) electrodeposited Co$_3$O$_4$ nanosheets formed thin film on a flexible Ti base via a one-step potentiostatic method, followed by an UV-ozone treatment for 30 min. The morphology of Co$_3$O$_4$ nanosheets the can be inherited by the sheet-like structure Co(OH)$_2$ precursor during the UV-ozone treatment. After optimized the electrodeposition time and NaOH concentration in the electrolyte, the Co$_3$O$_4$ modified Ti electrode (Co$_3$O$_4$/Ti) showed a much higher capacitance than that of the

Co(OH)$_2$ modified Ti electrode when used for a supercapacitor. A remarkably high specific capacitance of 1033.3 F g$^{-1}$ was achieved for the Co$_3$O$_4$ thin film at a charge-discharge current density of 2.5 A g$^{-1}$. The capacitance showed a high retention of 77% after the long-term stability test for 3000 repeated charge-discharge cycles. Another 2D Co$_3$O$_4$ structure is provided by Meher and Rao (Meher and Rao 2011). They synthesized highly porous ultra layered Co$_3$O$_4$ structures by hydrothermal method with simple homogeneous precipitation procedure. Micron size rectangular 2D flakes were well-arranged layer-by-layer to form the superstructures, showing large specific surface area, high pore volume, and uniform pore size distribution. The selective adsorption of the special molecule on the certain crystal planes will promote oriented growth of materials with a specific shape. Triton X-100 was used as a neutral surfactant to selectively adsorbed onto a highly exposed plan of the layered structure and prevented the agglomeration of 2D sheets. The formation of porous Co$_3$O$_4$ layered structure overall contained several intermediate processes including coalescence, hydrothermal, and Ostwald ripening, self-assembling, and thermal decomposition. The ultralayered Co$_3$O$_4$ demonstrated a high specific capacitance of 548 F g$^{-1}$ with ~100% Coulombic effeciency at a current density of 8 A g$^{-1}$ and maintained 66% of capacitance at 32 A g$^{-1}$.

In order to improve the electrochemical performance of Co$_3$O$_4$, hollow structured materials have been wildly studied. Co$_3$O$_4$ nanotubes (**Figure 16a**) have been successfully prepared via chemical deposition of cobalt hydroxide in anodic aluminum oxide (AAO) templates and calcining at 500 °C (Xu et al. 2010). The nanotube materials showed a specific capacitance of 574 F g$^{-1}$ at a current density of 0.1 A g$^{-1}$ and good capacitance retention of ~95% after 1000 cycles. Du et al. (Du et al. 2013) used cobalt acetate hydroxide (Co$_5$(OH)$_2$(CH$_3$COO)$_8$·2H$_2$O) prisms as the precursor and transferred them to cobalt oxide by calcination in the air. The solid 1D cobalt acetate hydroxide turned to be boxes with the hollow structure inside by the Kirkendall effect (**Figure 16b**). The Co$_3$O$_4$ boxes showed a higher capacitance than hexagonal Co$_3$O$_4$ nanosheets with similar specific surface area. 3D-nanonet hollow structured Co$_3$O$_4$ has been fabricated by calcination of a 3D-nanonet structured cobalt-basic-carbonate

precursor in the air without change of the original frame structure (**Figure 16c**) (Wang, Lei, et al. 2014). The precursor was synthesized by a heterogeneous precipitation process; the supposed formation mechanism is shown in **Figure 16d**. At the initial formation stage of the precursor nanocrystals, the $CO_2$ gas bubbles generated during the reaction could be a template for the aggregation of the nanocrystals with the help of the glucose. The 3D-nanonet hollow structured $Co_3O_4$ with a high specific surface area of 92.8 $m^2$ $g^{-1}$ delivered good rate capability and cyclic stability. Wang et al. (Wang, Pan, et al. 2014) fabricated a novel nanorod assembled multi-shelled cobalt oxide hollow microspheres (HSs) as shown in **Figure 16e,f**. This multi-shelled hollow structure was formed by calcination of a core-shell structured carbon microspheres (CS)@$Co_2CO_3(OH)_2$ composite which synthesized by the $Co_2CO_3(OH)_2$ nanorods vertically grown on the CS. When applied for supercapacitors, the multi-shelled $Co_3O_4$ hollow microspheres exhibited high capacitances of 394.4 and 360 F $g^{-1}$ at the current densities of 2 A $g^{-1}$ and 10 A $g^{-1}$, respectively. Besides the hollow structure design, a new attempt has been adopted. For example, mesoporous $Co_3O_4$ nanowires have been treated with $NaBH_4$ to obtain reduced $Co_3O_4$ nanowires, which can be used for oxygen evolution reaction (OER) and supercapacitors (**Figure 16g,h**) (Wang, Zhou, et al. 2014). The mesoporous $Co_3O_4$ nanowires with large surface area, 1D morphology for charge transport, and the low formation energy allow for efficient reduction treatment on the surface. Electrochemical supercapacitors based on the reduced $Co_3O_4$ nanowires revealed a much enhanced capacitance of 978 F $g^{-1}$ as compared with pristine $Co_3O_4$ nanowires (288 F $g^{-1}$) and decreased charge transfer resistance (**Figure 16i**). Density-functional theory (DFT) calculations indicated the formation of new gap states caused by oxygen vacancies. The electrons in the reduced $Co_3O_4$ that was previously related to the Co-O bonds inclined to be delocalized, leading to a much higher electrical conductivity (**Figure 16j**).

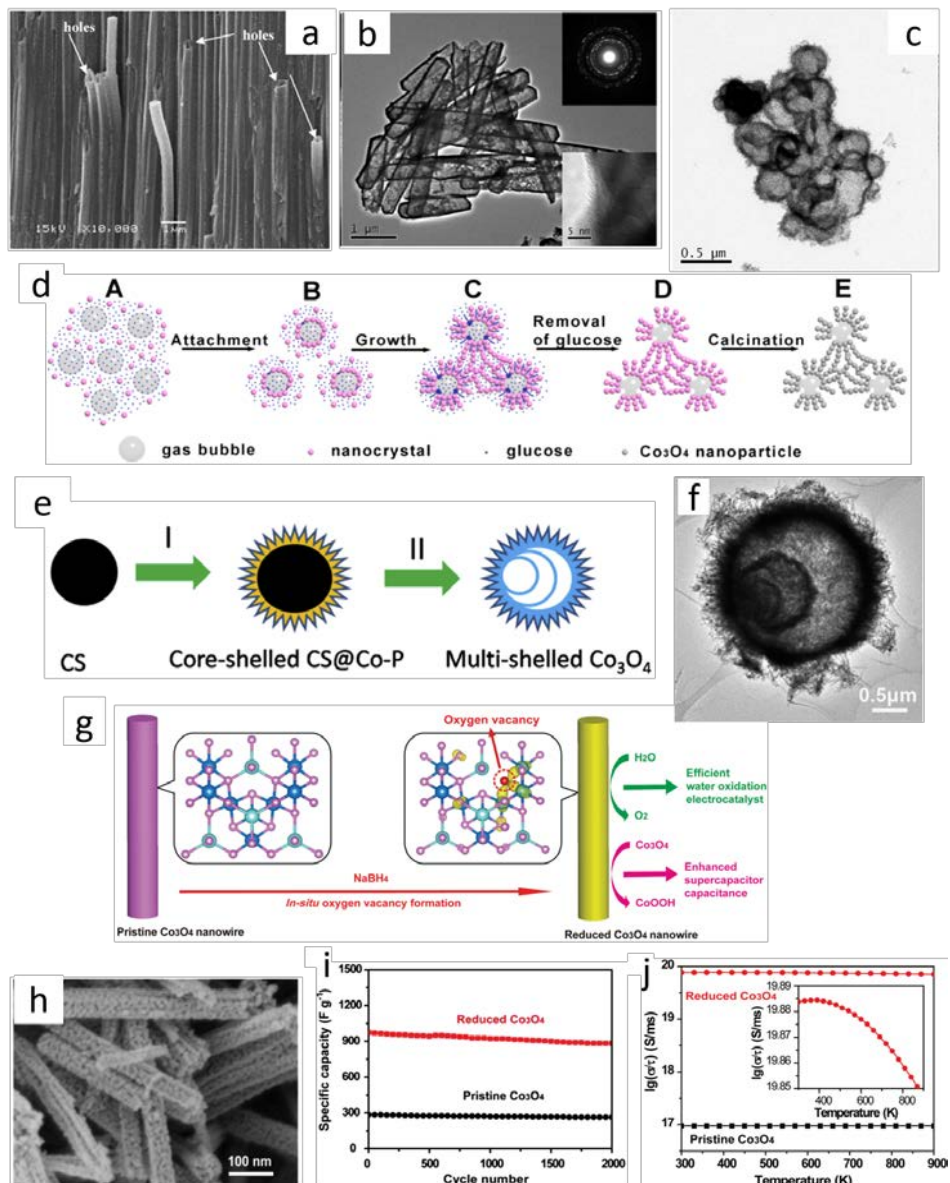

**Figure 16.** (a) SEM image of $Co_3O_4$ nanotubes. Reproduced with permission (Xu et al. 2010). Copyright 2010, Elsevier. (b) TEM image of $Co_3O_4$ boxes. Reproduced with permission (Du et al. 2013). Copyright 2013, Elsevier. (c) TEM image of 3D-nanonet hollow structured $Co_3O_4$. (d) Schematic representation of the possible formation mechanism of 3D-nanonet hollow structured $Co_3O_4$. Reproduced with permission (Wang, Lei, et al. 2014). Copyright 2014, American Chemical Society. (e) Schematic illustration of the preparation of core-shelled carbon micro-spheres@cobalt precursor (CS@Co-P) (step I) and their conversion to the multi-shelled $Co_3O_4$ hollow spheres (step II). (f) TEM image of the nanorod-assembled multi-shelled $Co_3O_4$ hollow microspheres. Reproduced with permission (Wang, Pan, et al. 2014). Copyright 2014, Elsevier. (g) Schematic of the $NaBH_4$ reduction for in situ creation of oxygen vacancies in $Co_3O_4$ NWs for efficient catalysis of oxygen evolution reaction and enhanced supercapacitor capacitance. (h) SEM image of the reduced $Co_3O_4$ nanowires. (i) Cycling stability of the reduced $Co_3O_4$ and pristine $Co_3O_4$ NWs tested at a current density of 2 A $g^{-1}$. (j) Calculated conductivity for the reduced $Co_3O_4$ and the pristine $Co_3O_4$. Inset: an expansion of the plot of reduced $Co_3O_4$. Reproduced with permission (Wang, Zhou, et al. 2014). Copyright 2014, Wiley.

**3.3.3** *Copper oxides*

Copper oxide, CuO, crystallizes with a monoclinic unit cell (JCPDS card no. 41-0254), with a crystallographic point group of 2/*m* or *C*2*h* (Xia, Zhang, et al. 2014). The space group of CuO unit cell is *C*2/*c*. CuO has been considered as a promising candidate for the electrode of supercapacitors due to its low cost, abundance, non-toxicity, and ease of preparation in diverse shapes of nanometer dimensions, and good electrochemical performance (Deng et al. 2014; Moosavifard et al. 2015; Mishra et al. 2018). According to the reaction mechanism, CuO is belonged to battery-type materials. The redox reactions involved in the transform between Cu(I) and Cu(II) species (Lin et al. 2011; Zhao et al. 2013; Zhang, Shi, et al. 2011):

$$CuO + 1/2H_2O + e^- \leftrightarrow 1/2Cu_2O + OH^- \tag{18}$$

or

$$CuO + H_2O + e^- \leftrightarrow CuOOH + OH^- \tag{19}$$

The theoretical pseudocapacitance of CuO can reach as high as ~1800 F g$^{-1}$ (Vidhyadharan et al. 2014), which is close to that of the widely studied hydrated ruthenium oxide (RuO$_2$·$n$H$_2$O) (~2200 F g$^{-1}$). However, most of the CuO nanostructures only achieve 15%-32% of the theoretical capacitance in the observed experiments. For example, CuO nanowires synthesized by chemical transformation approach only delivered a specific capacitance of 118 F g$^{-1}$ at the current density of 1 A g$^{-1}$ (Chen and Xue 2013). A leaf-like porous CuO-graphene nanostructure prepared by hydrothermal method could show an improved capacitance of 331.9 F g$^{-1}$ at 0.6 A g$^{-1}$ (Zhao et al. 2013). A hierarchical porous flower-shape CuO nanostructure composed of ultrathin nanoleaves with the large surface area and pore volume exhibited a remarkable specific capacitance of 520 F g$^{-1}$ at 1 A g$^{-1}$ (Lu et al. 2015).

Controlling the fabrication of hierarchical multifunctional architectures is a fascinating technique to enhance the electrochemical performance of metal oxides. Zhang et al. investigated the influence of the oriented assembly of low-dimensional building blocks on the performance of CuO (Zhang, Feng, et al. 2016). They synthesized two 3D CuO ordered nanostructures (CONs) assembled by 1D and 2D building blocks and compared them with disorganized 2D nanoflakes. The formation mechanisms were shown in **Figure 17a**. During the reaction, n-Butylamine was introduced as both the alkaline reagent and capping agent. The growth direction of the initial building blocks was guided by *n*-Butylamine. The different synthesis conditions would induce different oriented attachment mode of the initial building blocks and the assembly of the sub-nanostructure. The as-prepared CuO nanourchins and nanoflowers which assembled by 1D nanorods and 2D nanosheets delivered high specific capacitance (541 and 585 F g$^{-1}$ at 1 A g$^{-1}$), good rate capability (retaining 81% and 79% at 20 A g$^{-1}$), and stable cycle life (85.3% and 86.8% capacitance retention after 8000 cycles) as shown in **Figure 17b,c**. In contrast, disorganized 2D CuO nanoflakes only showed a low capacitance of 347 F g$^{-1}$ at 1 A g$^{-1}$. After this work, Zhang et al. (Zhang, Cui, et al. 2018) designed a uniform nanoporous CuO mesocrystal structure by a similar hydrothermal method at low-temperature. A similar reaction mechanism involved the adsorption of small *n*-butylamine molecules is demonstrated in **Figure 17d**. The CuO mesocrystals, which are different from the single crystals and polycrystals, were assembled by the oriented arrangement of CuO subunits (**Figure 17e**). The CuO mesocrystals exhibited single-crystal-like diffraction spots as shown in **Figure 17f**, indicating the high crystallinity and little lattice mismatch of mesocrystals, which supply high electronic conductivity and structural stability for electrode materials. Additionally, the porous structure provided numerous active sites for electrochemical reaction and fast ionic mobility. As a result, the CuO mesocrystals delivered a high pseudocapacitance of 612 F g$^{-1}$ at 1 A g$^{-1}$, which was much higher than that of CuO single crystals and polycrystals (**Figure 17g**).

However, it should be pointed out that CuO is not such suitable as good active materials for supercapacitors compared to other metal oxides such as $Co_3O_4$ and NiO because of its poor electrochemical redox reactivity and low discharge voltage in alkaline electrolyte.

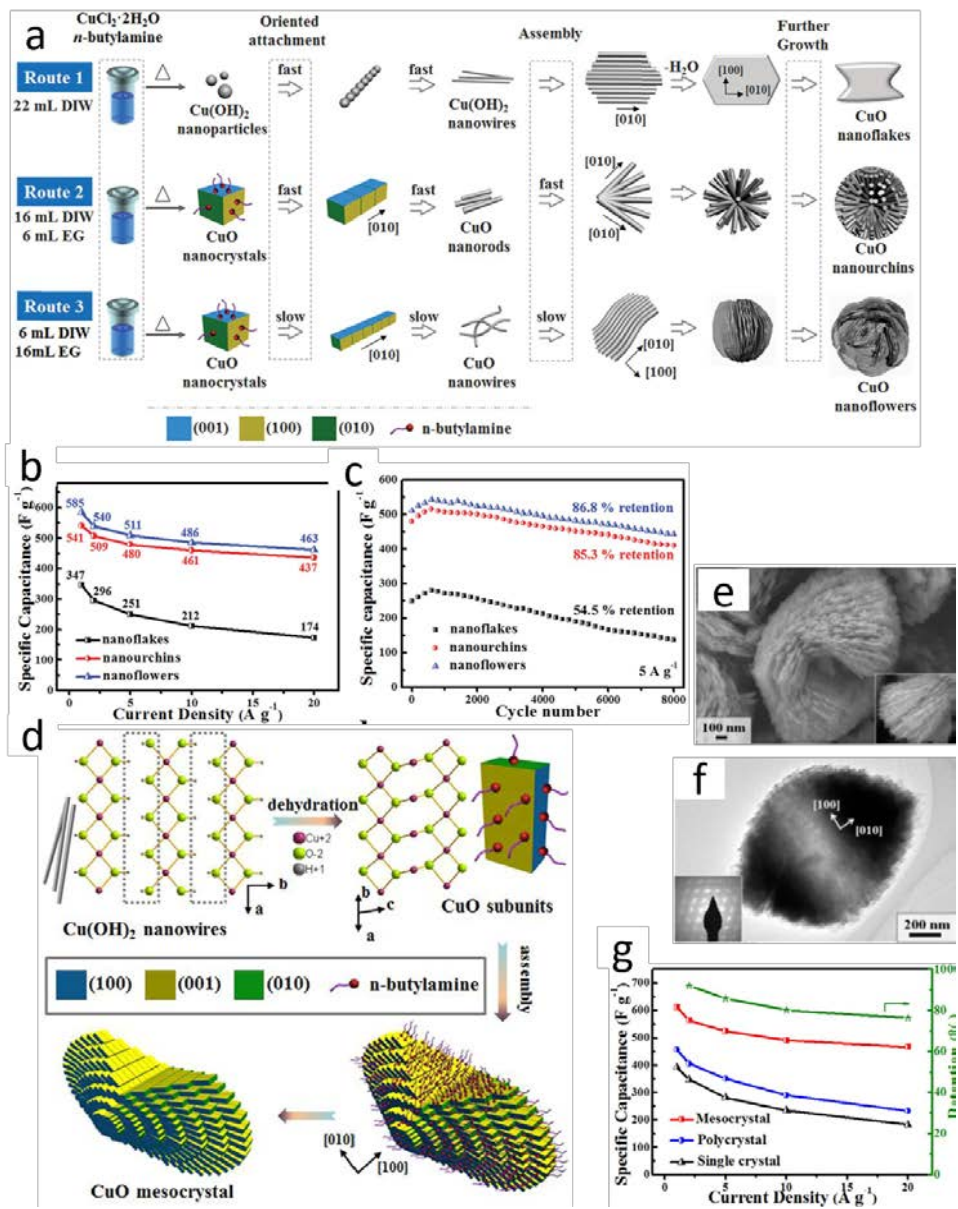

**Figure 17.** (a) Schematic illustration of the proposed mechanism for the formation of CuO architectures with various morphologies in different solvents. The rate performances of different CuO nanostructures in the range of current density from 1 to 20 A g$^{-1}$. (b) The cycling performances of different CuO nanostructures with 8000 cycling number at a current density of 5 A g$^{-1}$. Reproduced with permission (Zhang, Feng, et al. 2016). Copyright 2016, Royal of Society of Chemistry. (d) Schematic illustration of the proposed mechanism for the formation of CuO mesocrystals in the presence of n-butylamine. (e) SEM and (f) TEM images of CuO mesocrystals. The inset in panel (f) shows the corresponding SAED pattern. (g) Rate performances of different CuO structures from 1 to 20 A g$^{-1}$. Reproduced with permission (Zhang, Cui, et al. 2018). Copyright 2018, Elsevier.

### 3.3.4 *Binary TMOs*

Binary TMOs contain one transition metal ion and one electrochemically active or inactive metal ions. Binary TMOs have various crystal structures including spinel, scheelite, $CaFe_2O_4$-type and so on. Binary TMOs commonly exhibit much higher electric conductivity and electrocapacitive activity than single metal oxides and therefore have been widely used as electrode materials for supercapacitors. Spinel cobaltite ($MCo_2O_4$; M = Mn, Ni, Zn, Cu, etc.) is a promising candidate for supercapacitors due to its super high specific capacitance and good rate capability (Wang, Gao, et al. 2011; Wei et al. 2010). Moreover, spinel cobaltite combines the merits of both metal ions (Sharma et al. 2007). Among the various spinel cobaltites, $NiCo_2O_4$ is the most famous pseudocapacitance material. The spinel structure of $NiCo_2O_4$ is shown in **Figure 18a**, both Ni and Co ions have mixed valence states of +2 and +3, so the accepted general formula of the unique cobaltite is $Co_{1-x}^{2+}Co_x^{3+}(Co^{3+}Ni_x^{2+}Ni_{1-x}^{3+})O_4$ ($0 < x < 1$), where Ni cations occupy the octahedral sites and Co cations on both octahedral and tetrahedral sites (Wang, Liu, et al. 2012). The multivalences of transition metal cation lead to the transport of electrons between cations with relatively low activation energy, hence the $NiCo_2O_4$ has a relatively high electrical conductivity which is at least two orders of magnitude higher than single metal oxides of NiO and $Co_3O_4$ (Wang, Gao, et al. 2011; Wei et al. 2010). Additionally, the coexistence of Ni and Co improves the $NiCo_2O_4$ electrode with higher electrochemical activity. The charge storage reactions of $NiCo_2O_4$ in alkaline during the redox process are based on the following equations (Wang, Han, et al. 2012):

$NiCo_2O_4 + OH^- + H_2O \leftrightarrow NiOOH + 2CoOOH + 2e^-$ 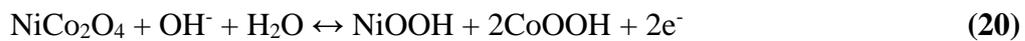 (20)

$CoOOH + OH^- \leftrightarrow CoO_2 + H_2O + e^-$ 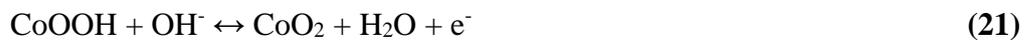 (21)

The development of $NiCo_2O_4$ nano-architecture can be classified into two different ways: (a) $NiCo_2O_4$ powder based materials and (b) $NiCo_2O_4$ conducting support based electrodes. So far, various nanostructures of $NiCo_2O_4$ powder have been reported for the electrochemical capacitor, such as nanoparticles (Wei et al. 2010; Hu et al. 2013), nanowires (Wang, Gao, et al. 2011; Yuan et al. 2012), nanosheets (Lu, Huang, et al. 2012), nanoflakes (Shakir et al. 2013), ordered mesoporous structure (Lu, Chen, Li, et al. 2013), nanoflowers (Zhang, Kuila, et al. 2015), nanoplates (Pu et al. 2013), and urchin-like nanostructures (Xiao and Yang 2011; Wang, Liu, et al. 2012). Recently, conducting substrate supported $NiCo_2O_4$ has received extensive attention due to the enhanced electronic conductivity by growing thin layer of nanostructured $NiCo_2O_4$ on the surface of a porous, high surface area and electronically conducting substrates. Moreover, the concept of the binder-free and free-standing electrode with the conductive substrates such as Ni-foam, carbon fiber paper (CFP), flexible carbon fabric, etc. makes $NiCo_2O_4$ as a popular research material. Xiong Wen (David) Lou's group has reported many important works on this. For instance, they prepared mesoporous $NiCo_2O_4$ nanosheets on four different types of conductive substrates, including Ni foam, Ti foil, stainless-steel foil, and flexible graphite paper by a chemical bath deposition method (**Figure 18b,c**) (Zhang and Lou 2013). They found the $NiCo_2O_4$ nanosheets-Ni foam deliver the best electrochemical performance, maintaining a high areal capacitance of 1.95 F cm$^{-2}$ over 3000 cycles at a current density of 8.5 mA cm$^{-2}$ (**Figure 18d**). Besides nanosheets, other nanostructure array on various substrate have been reported to show excellent pseudocapacitance performance, such as nanowires/carbon textiles (Shen et al. 2014), nanoneedle arrays/Ni foam and Ti foil (Zhang et al. 2012), and core/shell nanoflake array/Ni foam (Liu, Shi, et al. 2013).

Other mixed metal oxide, such as $MMoO_4$ (M = Ni, Co) (Senthilkumar et al. 2013; Liu et al. 2012), $MFe_2O_4$ (M = Ni, Co, Sn, Mn) (Kuo and Wu 2005), $MSnO_3$ (M = Ni, Co) (Saranya and Selladurai 2018), and $NiMn_2O_4$ (Zhang, Guo, et al. 2013) has also been applied in energy storage devices.

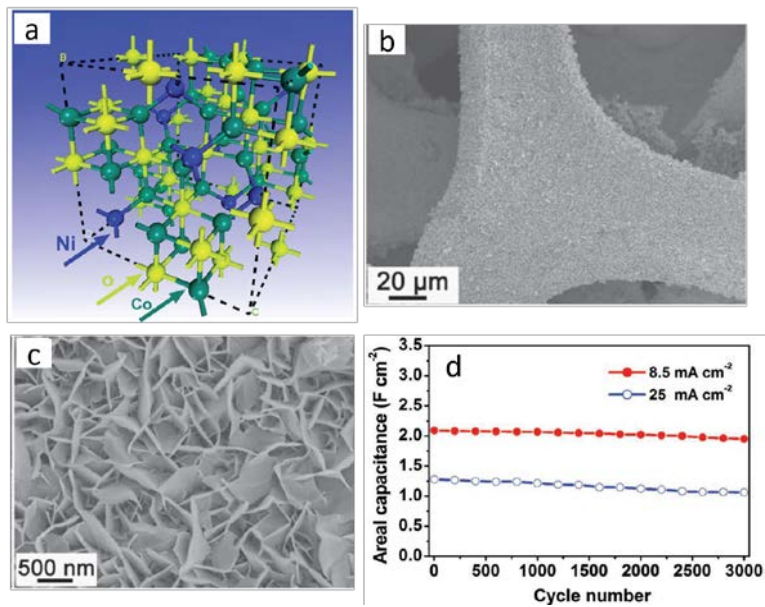

**Figure 18.** (a) Crystal structure of NiCo$_2$O$_4$ unit cell with the spinel structure. Reproduced with permission (Wang, Liu, et al. 2012). Copyright 2012, Royal of Society of Chemistry. (b, c) SEM images of NiCo$_2$O$_4$ nanosheets on Ni foam. (d) The capacitance as a function of cycle number at constant current densities of 8.5 and 25 mA cm$^{-2}$. Reproduced with permission (Zhang and Lou 2013). Copyright 2013, Wiley.

**4. TMOs-based composites for SCs**

Although nanostructure engineering has been confirmed to be a useful strategy in enhancing the electrochemical performance (including the reversible capacity/capacitance, cycling stability, and rate performance) for TMO-based electrodes compared with their bulk counterparts, single component materials that are achieved by simply engineering the nanostructure (tuning the particle size, geometric shape, and porosity) is still unsatisfactory in terms of ultra-stable, high energy density and high power density supercapacitors for TMO-based electrode materials. Composite materials have been explored to improve the electrocapacitive performance of SCs, which are formed by a TMO material and another material with complementary properties, or a material that can generate synergetic effects with the host material.

The composite materials can be derived into two types according to the reaction mechanism. The initial one is to combine the TMO material with carbon-based materials or conductive polymers. The TMO materials can grow on the carbon-based materials or conductive polymers with the support of them. This kind of composite can increase the specific surface area, buffer the volume change, decrease the agglomeration and boost the electric conductivity. Another type of composite materials is known as the "hierarchical structure", which contains one TMO material and one metal-based material. The two components in the hierarchical structure usually show complementary properties, resulting in an enhanced synergetic effect. Furthermore, the hierarchical structure can well prevent the aggregation of electroactive materials even when the particle size and mass loading increase. The hierarchical structure is particularly suitable for practical application to achieve high energy density because both of the components exhibit high capacitance.

**4.1** Carbon-based composites

The storage mechanism of carbon materials is the electrical double layer capacitors (EDLCs), which can store energy by pure electrostatic attraction between electrolyte ions and charged electrode surfaces, without involving chemical reactions during charging/discharging processes (Xiong et al. 2015). This storage mechanism provides fast charging/discharging processes leading to high power density and superb cycling stability. Moreover, many outstanding properties, such as high surface area, lightweight, good electrical conductivity, controlled pore size distribution and compatibility with other materials, make the carbon materials as promising pairing materials for TMO composite.

**4.1.1** *TMOs/CNTs*

Carbon nanotubes (CNTs), discovered in 1991, are cylindrical nanostructure wrapped by graphene sheets that are covalently bonded through $sp^2$ hybridization. Although CNTs are a prominent SC electrode material, they suffer from relatively low surface area and low density, leading to difficulty in their assembly into electrodes by using traditional electrode fabrication methods compared with other advanced carbon materials. TMO/CNTs composite shows a promising way for practical applications due to the high energy density of TMO. CNTs have a unique internal structure, their open-ended tunnel is beneficial for accelerating the diffusion of ions to the active surfaces of the composites. Additionally, CNTs show excellent physical and chemical stability, low gravimetric density and excellent electronic conductivity (105 S cm$^{-1}$), which makes the CNTs an excellent mesoporous matrix for TMO materials.

The pseudocapacitive material $\alpha$-Fe$_2$O$_3$ nanohorns was directly deposited into the 3D porous sponge by low-temperature hydrothermal synthesis (Cheng, Gui, et al. 2015). The original interconnected CNT network and high conductivity can be protected during synthesis, resulting in stable electrochemical characteristics under high compression. Bai et al. (Bai et al. 2015) synthesized homogeneous ZnCo$_2$O$_4$ nanoflowers on a 3D layered structure of CNTs/nitrogen-doped graphene film via a hydrothermal process and subsequent calcination method, which display a high specific capacitance of 1802 F g$^{-1}$ at 1 A g$^{-1}$ and good cycling stability. However, the large size of ZnCo$_2$O$_4$ nanoflowers (the average diameter is 4 µm) obstruct the application of them for the high-performance SCs. Sun et al. (Sun, Deng, et al. 2015) evaporated a thin layer of WO$_3$ on a free-standing CNT film flexible negative electrode via a facile vacuum filtration and a subsequent physical vapor deposition method, and fabricated asymmetric supercapacitors (ASCs) with pure CNT films as the paired positive electrodes. Although the volumetric capacitance of this electrode reaches a high value of 2.6 F cm$^{-1}$, the separate layers of WO$_3$ and CNT film can't play a synergistic role for each other. As a result, only 75.8% of the capacitance of the ASC can be maintained after 50000 cycles. Therefore, how to obtained homogeneous and closely integrated MO/CNTs composite is a big challenge.

The dispersion of 0D MO nanoparticles onto CNTs is a general way to suppress the reunion of the MO compounds. For example, Guan et al. (Guan, Liu, et al. 2015) synthesized a novel hierarchical structure composed of iron oxide nanoparticles decorated on a 3D ultrathin graphite foam-carbon nanotube forest substrate (noted as GF-CNT@Fe$_2$O$_3$) by a CVD method for carbon materials and an ALD technique for metal oxide. As shown in **Figure 19**, the nano-crystalline Fe$_2$O$_3$ has been uniformly deposited on the structure of graphite foam with CNT forest. GF-CNT@Fe$_2$O$_3$ shows a high areal capacitance of ~470.5 mF cm$^{-2}$, which is ~4 times larger than that of GF-CNT (~93.8 mF cm$^{-2}$) at the same current density of 20 mA cm$^{-2}$. 72.4% of the capacitance is maintained when the current density increased from 5 to 40 mA cm$^{-2}$. The capacitance loss is only 6.5% for GF-CNT@Fe$_2$O$_3$ after 50000 cycles. The excellent electrochemical performance can be attributed to high mechanical stability of the structure, the small nanocrystallite of Fe$_2$O$_3$, and the uniform distribution of Fe$_2$O$_3$ particles. Besides Fe$_2$O$_3$, confined NiO nanoparticles were also successfully deposited uniformly on the GF-CNT by the same method (Guan, Wang, et al. 2015). Then, an improved ALD technique was investigated to synthesize NiO nanoparticles coated CNTs by simultaneously employing O$_3$ and H$_2$O as oxidizing agents in a single ALD cycle of NiO for the first time. The capacitance of this composite is 622 F g$^{-1}$ at 2 A g$^{-1}$ and 2013 F g$^{-1}$ for NiO, with 26% of capacitance loss at 50 A g$^{-1}$ (Yu, Wang, et al. 2016).

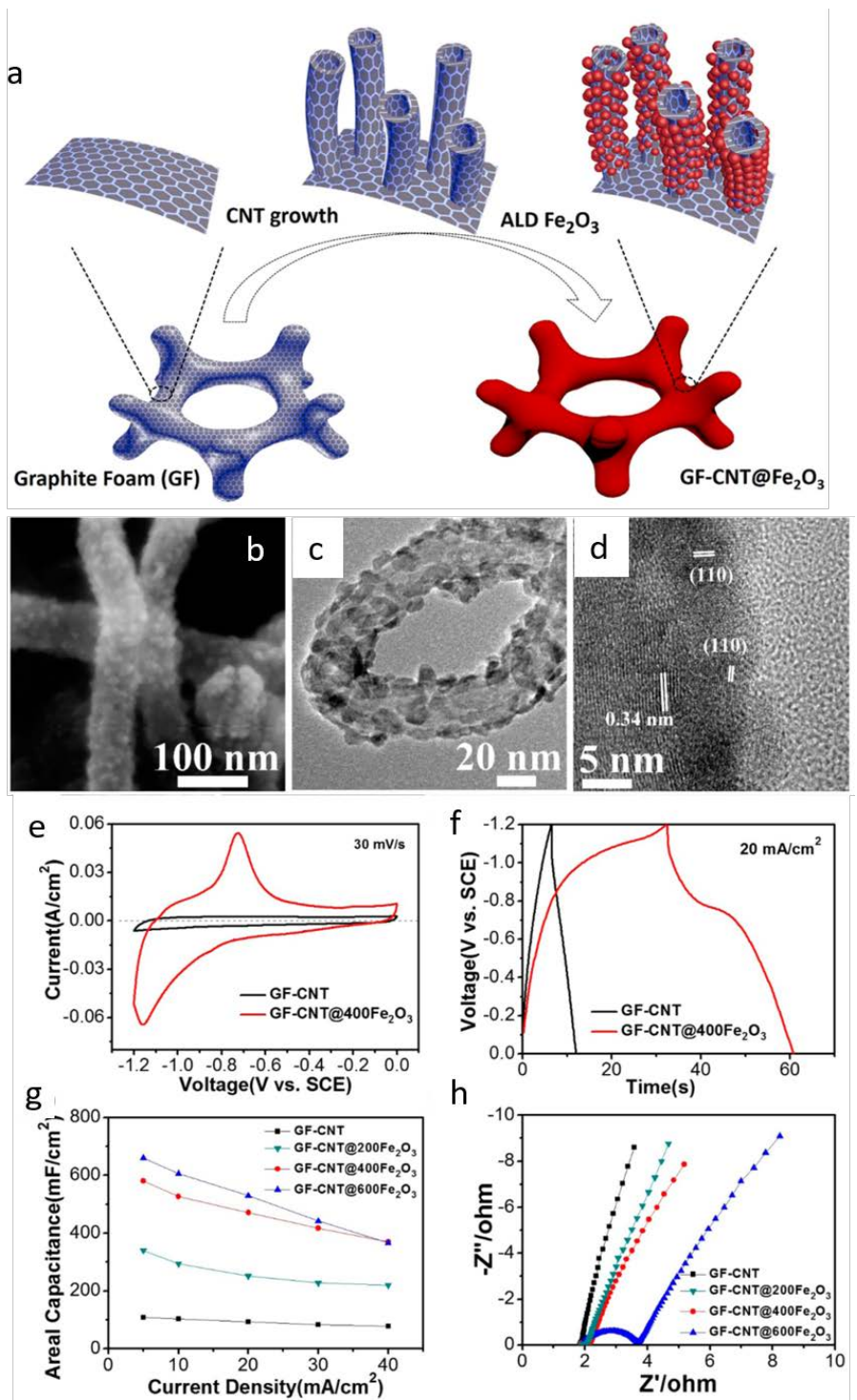

**Figure 19.** (a) Growth Procedure of GF-CNT@Fe$_2$O$_3$ Starting from Graphite Foam. (b) SEM and (c, d) HRTEM images of GF-CNT@Fe$_2$O$_3$. (e) CV curves and (f) charge-discharge curves of GF-CNT@400Fe$_2$O$_3$ and GF-CNT. (g) Rate properties and (h) EIS of four samples. Reproduced with permission (Guan, Liu, et al. 2015). Copyright 2015, American Chemical Society.

Another effective approach for enhancing the performance is to tailor the internal structure of TMO active materials. Hollow structured RuO$_2$ nanoparticles/CNTs composite was prepared by a core-shell templated approach involving the loading of core-shell Ag-Ru nanoparticles on carbon nanotube support and follow-up removal of the Ag component (Wang, Xu, et al. 2015). The outstanding electrochemical performance (including high specific capacitances of 819.9 F g$^{-1}$, good rate, and cycling performance) can be attributed to the unique hollow structure and the supported CNT, preventing $RuO_2$ from aggregation. Many other TMO nanoparticles such as $V_2O_5$ (Yilmaz et al. 2016), ZnO (Ramli et al. 2017), $MnO_2$ (Huang, Zhang, et al. 2015), $RuO_2$ (Kong et al. 2017), NiO (Nunes et al. 2019), $NiFe_2O_4$ (Kumar et al. 2018), $ZnFe_2O_4$ (Raut et al. 2018), $MFe_2O_4$, (M(II) = Mn, Fe, Co) (Pereira et al. 2018) have been wildly used for MO/CNTs composite.

Recently, many multi-component active materials have been applied to form CNTs hybrid. For instance, $SnO_2$-CuO-$Cu_2O$ ($SnO_2$-$Cu_xO$) porous thin films on CNTs core-shell structure was synthesized through an electroless deposition method by using CNTs as the substrate (Daneshvar et al. 2018). The high specific surface area, small particle size, the synergistic effect of Sn, and conductivity improvement by using CNTs enable an excellent performance with a high specific capacity of 662 F g$^{-1}$. Other multi-component materials such as NiO/Ni(OH)$_2$ nanostructures (Palani et al. 2018) and Co-$Co_3O_4$ (Zou et al. 2018) are also explored for CNT hybrid.

In addition to 0D nanoparticles, 2D nanosheets of MO compounds are also used to form a 1D-2D hierarchical structure. A typical carbon nanotubes@nickel oxide nanosheets (CNT@NiO) core-shell composites were synthesized through a facile chemical bath deposition method followed by thermal annealing (Yi et al. 2015). The interconnected NiO nanosheets are branching uniformly on the individual CNT surface, creating a highly porous network, which can provide a high surface area and more active sites contacted with electrolyte ions, and consequently result in high utilization of the NiO shell. The CNT@NiO displays high capacitance of 996 F g$^{-1}$ at 1 A g$^{-1}$ within 0-0.5 V. The asymmetric supercapacitors coupled with porous carbon polyhedrons (PCPs) as negative electrode shows maximum energy density of 25.4 Wh kg$^{-1}$ at a power density of 400 W kg$^{-1}$ in a broad potential range of 0-1.6 V. Nan and partners (Nan et al. 2015) designed a novel nanocomposite consisting of bimetal oxide $NiMn_2O_4$ nanosheets with CNTs. The nanosheets are in mutual contact and are uniformly grown on the surface of the CNTs backbone, which means that CNTs are highly suitable substrates for spinel nickel manganese's nucleation and crystal growth. Brown et al. (Brown et al. 2015) electrodeposited $MnO_2$ nanosheets on graphenated carbon nanotubes (g-CNTs), finding that thin and conformal sheets can be deposited on g-CNTs while thick layer or "block" of $MnO_2$ are obtained on CNTs. Then they investigated the influence of the mass loading on the morphology and performance. When the mass loading is no greater than 2.3 mg cm$^{-2}$, the nanosheets are uniformly dispersed, the capacitance is increasing with the increase of loading; when the loading is larger than 2.3 mg cm$^{-2}$, it shows thick $MnO_2$ film deposition, and the capacitance decreases significantly.

Another application aspect of TMO/CNTs composite is to fabricate the flexible binder-free and free-standing electrode without a current collector. This kind of electrode for SCs shows improved ionic and electronic transfer. For example, $V_2O_5$/CNT free-standing films prepared by a continuous blade coating process greatly facilitates electronic transport along CNTs while maintaining rapid ion diffusion within $V_2O_5$ nanowires and fast electron transfer across the $V_2O_5$/CNTs interfaces (Wu, Gao, et al. 2016). Lee et al. (Lee et al. 2018) employed $MoO_3$ and $MnO_2$ for the two electrodes of asymmetric SCs due to the broad potential window up to 2V for the two materials. The free-standing electrodes were formed by the nanostructured materials with CNTs, leading to $MoO_3$ nanodots on CNT surfaces ($MoO_3$@CNT) and mesoporous $MnO_2$ embedded in CNT networks ($MnO_2$@CNT). These electrodes showed high specific surface area and electric conductivity (68 m$^2$ g$^{-1}$ and 2.27 S cm$^{-1}$ for $MoO_3$@CNT, 343 m$^2$ g$^{-1}$ and 10.82 S cm$^{-1}$ for $MnO_2$@CNT). The assembled asymmetric supercapacitors exhibited high power (10000 W kg$^{-1}$) and high energy (27.8 Wh kg$^{-1}$). The fiber-shaped SC is a promising energy storage device for wearable and portable electronics. TMO/CNTs hybrid has been developed for fiber device due to the superb electrical conductivity, flexibility, and mechanical strength of CNT. Su and cooperators (Su et al. 2016) filament spun thin stainless steel (SS) into the center of a CNT yarn, followed by the modification of $MnO_x$ or $Co_3O_4$ via electrochemical deposition method. The as-prepared MnO$_x$/CNT/SS and Co$_3$O$_4$/CNT/SS delivered good electrochemical properties and high flexibility even under a sharp bending deformation. However, low capacitance and energy density is a big shortcoming for the fiber-shaped SCs because the two fiber electrodes always show low surface area and operating voltage range. For this, coaxial fiber-shaped asymmetric supercapacitors (CFASCs) composed of pseudocapacitive positive and negative electrodes have been prepared by wrapping a PVA-LiClO$_4$ gel electrolyte-coated Fe$_2$O$_3$/CFs core with MnO$_2$/CNT-web paper (Patil et al. 2018). The as-prepared MnO$_2$/CNT-web//PVA-LiClO$_4$//Fe$_2$O$_3$/CFs CFASC displayed an enhanced volumetric energy density of 0.43 mWh cm$^{-3}$ with good cycling stability, as well as outstanding flexibility.

### 4.1.2 *TMOs/graphene*

The structure of graphene is defined as the arrangement of sp$^2$-bonded carbon atoms in a honeycombed single layer. Graphene was firstly marked as an 'academic material', became a real 'miracle material' soon after it was successfully isolated in 2004. It should be noted that graphene is regarded as a single-layer (or monolayer) sheet of graphite at the initial stage of its development (Novoselov et al. 2004). In recent years, the term 'graphene' is frequently expanded to bi-/double-layer (Lee et al. 2014) and few-layer (Paton et al. 2014) where 'few' commonly refers to less than 10 layers including bilayers unless otherwise stated (Schedin et al. 2007). The term 'multilayer' (Güell et al. 2012), which is also used to define graphene, contains bilayer and few-layer but is not rigorously limited to below 10 layers.

The properties of graphene can be tuned within broad ranges through the arrangement of the atoms by various way although it contains only one element. Graphene has been widely explored as the most promising candidate of an ideal electrode material due to its excellent properties such as wide potential windows, high availability of functional groups with a large surface area, superior optical trenchancy, inherent flexibility, and excellent thermochemical stability, processability, and electrical performance (Pumera 2010; Yang et al. 2013; Yu et al. 2013). In particular, the specific surface area of graphene is up to 2630 m$^2$ g$^{-1}$, which is much higher than that of black carbon (< 900 m$^2$ g$^{-1}$) and CNTs (50-1315 m$^2$ g$^{-1}$) but similar to ACs. Most of the widely developed TMOs such as MnO$_2$, Co$_3$O$_4$, V$_2$O$_5$, etc. are always low electronic conductivity that is bandgap semiconductors or even insulators, which severely impede their practical application. Besides, single TMO-based SCs usually exhibit lower capacities and energy densities than their theoretical value in practice due to the relatively shallow surface redox reaction area (with a thickness of only a few nanometers) of TMOs. Therefore, the high specific surface area and high electrical conductivity (106 S m$^{-1}$) indicates that graphene is an ideal matrix to hybridize with TMO compounds. Lee et al. (Lee et al. 2015) used GO as substrate and formed graphene/VO$_2$ hybrid films. They found that the ultra-large graphene shows lower sheet resistance (0.57 kΩ/sq.) than small scale graphene (with a sheet resistance of 55.74 kΩ/sq.). Interestingly, the specific capacitances of the large and small graphene are 769 F g$^{-1}$ and 385 F g$^{-1}$ at the same current density. It means that graphene with higher conductivity indicates greater potential in hybrid electrodes.

Additionally, graphene is well compatible with dissimilar active components (e.g. metal oxides, transition metals and conducting polymers), thereby resulting in the formation of high-stable graphene-based composites with strong bindings by in-situ hybridization and ex-situ recombination (Huang et al. 2012). Many different nanostructures of active materials including anchored, wrapped, encapsulated, layered, sandwich-like and mixed modes can combine with graphene in the hybrid, forming 1D, 2D or 3D macroscopic architectures (Raccichini et al. 2015). There are four typical outstanding features of the TMO/graphene composite for SCs: (a) the nucleation and growth of TMO can be induced by 2D support of graphene, on which well-defined sizes, shapes and crystallinity of uniform TMOs can be formed; (b) the agglomeration of TMO compounds can be effectively suppressed by graphene. The contact of the electrode with the electrolyte is extended through the large accessible surface area, and the volume change of the TMO active materials during cycling process can be alleviated; (c) the TMO compounds always suffer from the poor conductivity which can be highly enhanced by graphene owing to its superb electrical conductivity; (d) the strong interfacial interactions between graphene and TMOs cause by oxygen-containing groups on the surface of graphene facilitate the electron transfer.

We can use the in-situ deposition method to homogenously locate nanostructured TMOs on graphene base. Take $RuO_2$ for example, although $RuO_2$ has a high theoretical capacitance (1358 F $g^{-1}$) and metallic conductivity (300 S $cm^{-1}$), $RuO_2$ nanoparticles easily suffer from large aggregates, inducing incomplete redox reactions and significant degradation in the electrochemical performance. Hence well dispersion of $RuO_2$ on graphene is significantly important. Hydrous $RuO_2$/graphene sheet hybrids can be synthesized by combining sol-gel and low-temperature annealing processes with RGO and $RuCl_3$ as feedstock (Wu, Wang, et al. 2010). With the help of oxygen-containing functional groups, 38.3 wt% Ru sample with enlarged SSA of 281 $m^2$ $g^{-1}$ showed fine $RuO_2$ particles (5-20 nm) anchored onto the surface of graphene sheets by investigating the influence of different Ru loadings. The $RuO_2$/RGO composite showed high specific capacitance of 570 F $g^{-1}$, high energy density (20.1 Wh $kg^{-1}$) and high power density (10000 W $kg^{-1}$). Yang et al. (Yang, Liang, et al. 2015) used GO/$RuCl_3$ aqueous dispersions to prepare graphene/$RuO_2$ hydrogels by hydrothermal method. $Ru^{3+}$ ions were uniformly adsorbed on GO sheets in the suspension because positively charged $Ru^{3+}$ ions and negatively charged GO layers show strong electrostatic interactions, resulting in homogeneous dispersion of $RuO_2$ nanoparticles on the exfoliated RGO sheets. Kim's group (Kim et al. 2013) revealed that the number density of $RuO_2$ nanoparticles decreases with the increase of the C/O ratio of reduced GO during the in situ chemical synthesis approach. That is because much more anchoring sites will be provided by the chemical interaction between the Ru ions and the oxygen-containing functional groups where the nucleation of particles takes place. In addition, the conductivity of RGO in $RuO_2$/RGO nanocomposites has been significantly improved by the microwave-hydrothermal process.

Zhang et al. (Zhang, Jiang, et al. 2011) explored a novel ASC by adopting RGO-$RuO_2$ as anode and RGO-PANi as cathode materials. For comparison, symmetric SCs of RGO-PANi//RGO-PANi and RGO-$RuO_2$//RGO-$RuO_2$ showed the specific capacitance of~400 and ~340 F $g^{-1}$, respectively, at 0.3 A $g^{-1}$. The capacitive performance of the ASC is highly enhanced compared with that of the symmetric SCs fabricated with RGO-RuO$_2$ or RGO-PANi as electrodes. An energy density of 26.3 W h kg$^{-1}$, was obtained for ASC due to the broadened potential window in an aqueous electrolyte, which is about two times higher than that of the symmetrical SCs based on RGO-RuO$_2$ (12.4 Wh kg$^{-1}$) and RGO-PANi (13.9 Wh kg$^{-1}$) electrodes. Moreover, the ASC showed a high power density of 49.8 kW kg$^{-1}$ at an energy density of 6.8 Wh kg$^{-1}$. Dai's group (Wang, Liang, et al. 2011) also assembled asymmetrical SC by pairing up a Ni(OH)$_2$/graphene electrode with a RuO$_2$/graphene electrode operating in aqueous solutions at a voltage of ~1.5 V, which displayed significantly higher energy densities than symmetrical RuO$_2$-RuO$_2$ SCs or asymmetrical SCs based on either RuO$_2$-carbon or Ni(OH)$_2$-carbon electrode pairs. The ASC showed high energy density of ~48 Wh kg$^{-1}$ at a power density of ~0.23 kW kg$^{-1}$, and high power density of ~21 kW kg$^{-1}$ at an energy density of ~14 Wh kg$^{-1}$. Recently, Kong and partners (Kong et al. 2017) designed a unique nanostructure electrode consisting of RuO$_2$ nanoparticles with ultra-fine diameter (1.9 nm) anchored on the surface of graphene nanosheets (GNS) and carbon nanotube (CNT). The as-prepared RuO$_2$-GNS-CNT electrode can be directly used as a 3D binder-free electrode, exhibiting a high capacitance of 480.3 F g$^{-1}$ at 0.6 A g$^{-1}$ and excellent long-term cyclic stability (89.4% capacitance retention over 10000 cycles). Additionally, the RuO$_2$-GNS-CNT based symmetry supercapacitors demonstrate a maximum energy density of 30.9 Wh kg$^{-1}$ and power density of 14000 W kg$^{-1}$.

Manganese oxide is one of the most promising positive electrode materials for SCs, unfortunately, its poor conductivity (10$^{-5}$-10$^{-6}$ S cm$^{-1}$) limits the rate capability for high-power applications. Highly conductive (5500 S m$^{-1}$) graphene networks were applied to improve the conductivity of MnO$_2$ by electrodeposition (He et al. 2012). A high area capacitance of 1.42 F cm$^{-2}$ (2 mV s$^{-1}$) and power density (62 W kg$^{-1}$) can be achieved by this graphene/MnO$_2$ composite based supercapacitors. N-doped graphene (NG) has been used as both the support of electrodeposition of MnO$_2$ electrodes and the negative electrode (Wang et al. 2019). The NG//MnO$_2$ asymmetric micro-supercapacitors showed a wild potential window (1.8 V) and a specific capacitance (13 mF cm$^{-2}$). Chen et al. (Chen et al. 2010) synthesized a composite of graphene oxide supported by needle-like MnO$_2$ nanocrystals (GO-MnO$_2$ nanocomposites) through a simple soft chemical route in a water-isopropyl alcohol system. The formation mechanism of the one-dimensional structure is shown in **Figure 20a,b**: firstly manganese ions intercalate into or adsorb onto the GO sheets, then the crystal species nucleate and grow in the double solvent system by dissolution-crystallization and oriented attachment mechanisms. Furthermore, the chemical interaction between GO and MnO$_2$ was found to increase the electrochemical performance of as-prepared nanocomposites. One-dimensional α-MnO$_2$ nanowires/graphene composite also can be fabricated by (MGC) by solution-phase assembly of graphene sheets and MnO$_2$ nanowires (Wu, Ren, et al. 2010). The ASC assembled by MGC and graphene electrodes exhibits superior energy density of 30.4 Wh kg$^{-1}$, high power density (5000 W kg$^{-1}$ at 7.0 Wh kg$^{-1}$) and acceptable cycling performance of 79% retention after 1000 cycles (**Figure 20c-f**). Cui's group (Yu et al. 2011) developed a "conductive wrapping" method to further increase the conductivity of MnO$_2$, and thus significantly enhance the supercapacitor performance of graphene/MnO$_2$-based nanostructured electrodes. As shown in **Figure 20g,h**, by the 3D conductive wrapping of graphene/MnO$_2$ nanostructures with carbon nanotubes or conducting polymer, the specific capacitance of the electrodes (considering total mass of active materials) has substantially increased by ∼20% and ∼45%, respectively, with values as high as ∼380 F g$^{-1}$ achieved.

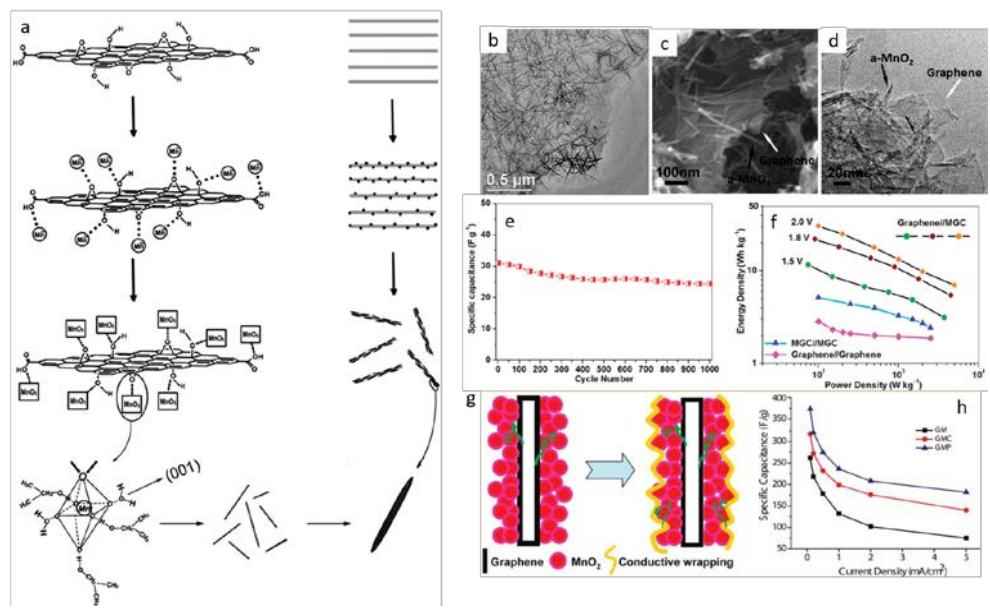

**Figure 20.** (a) The formation mechanism and (b) TEM image of GO–MnO$_2$ nanocomposites. Reproduced with permission (Chen et al. 2010). Copyright 2010, American Chemical Society. (c) High-magnification SEM images and (d) TEM image of MnO$_2$ nanowire/graphene composite (MGC). (e) Cycle performance of the graphene//MGC asymmetric EC with a voltage of 2.0 V at a current density of 500 mA g$^{-1}$. (f) Ragone plot related to energy and power densities of graphene//MGC asymmetric ECs with various voltage windows, graphene//graphene and MGC//MGC symmetric ECs. Reproduced with permission (Wu, Ren, et al. 2010). Copyright 2010, American Chemical Society. (g) Schematic illustration showing conductive wrapping of graphene/MnO$_2$ (GM) to introduce an additional electron transport path (in a discharge cycle). (h) Summary plot of specific capacitance values for three different electrode systems: GM-, GMC-, and GMP-based textiles at various current densities. Reproduced with permission (Yu et al. 2011). Copyright 2011, American Chemical Society.

The optimization of the active material on the graphene surface is a good strategy for further improvement of electrocapacitive performance of graphene-based composites. Multivalent manganese oxide Mn$_3$O$_4$/reduced graphene oxide (rGO) nanohybrid paper is fabricated through a simple gel formation and electrochemical reduction process by Hu et al. (Hu, Guan, et al. 2015) The confined Mn$_3$O$_4$ nanofibers well dispersed within the rGO sheets ensure high electrical conductivity and high mass loading of Mn$_3$O$_4$ nanofibers (0.71 g cm$^{-3}$) without sacrificing the capacitance. The Mn$_3$O$_4$/rGO hybrid paper-based ASC showed a specific capacitance of 54.6 F g$^{-1}$ at 1 mV s$^{-1}$ and energy density of 0.0055 Wh cm$^{-3}$, and good cycling stability (85% capacitance retention after 8000 cycles at 2 A g$^{-1}$). Wang et al. (Wang, Lai, et al. 2017) prepared MnO$_x$/rGO composite which was formed by the reduction of $\delta$-MnO$_2$ nanosheets on GO. The GO template keeps the MnO$_x$ nanoparticles uniformly dispersed on rGO conductive network. Due to the mixed-valence state Mn cations in MnO$_x$, abundant surface redox reaction sites were introduced, which showed synergistic effect with rGO. The rGO/MnO$_x$ composite exhibits a high specific capacitance of 202 F g$^{-1}$ and ultra-long cycling lifetime with 106% capacitance retention after 115000 cycles at 12 A g$^{-1}$. The MnO$_x$/rGO assembled ASC device exhibited considerable energy density of 49.7 Wh kg$^{-1}$ and excellent cycling stability of 96% capacitance retention after 80000 cycles at 5 A g$^{-1}$ in an ionic liquid electrolyte. Many other high-voltage aqueous ASC devices assembled by graphene/MnO$_2$ nanosheets (Fan et al. 2011; Xu, Wei, et al. 2015; Qu et al. 2009), graphene/MnO$_2$ films (Jin et al. 2013; Sumboja et al. 2013; Shao et al. 2013), or 3D graphene/$MnO_2$ frameworks (Wu, Winter, et al. 2012; Wu, Sun, et al. 2012) are well developed, which could operate at a large working voltage of ≥1.8 V, and greatly widen the maximum operating voltage of aqueous electrolytes.

In addition to $RuO_2$ and $MnO_x$, other TMO compounds have also been hybridized with graphene to form the composite materials, these hybrids including NiO (Zhao et al. 2011; Wu et al. 2014), $Fe_3O_4$ (Qu et al. 2011; Qi et al. 2013), $Co_3O_4$ (Yan et al. 2010; Xiang et al. 2013), $V_2O_5$ (Foo et al. 2014; Wu et al. 2015; Perera et al. 2013), ZnO (Haldorai et al. 2014), $SnO_2$ (Chen et al. 2014), $CeO_2$ (Wang, Guo, et al. 2011), $MoO_3$ (Chang et al. 2013) as well as $NiCo_2O_4$ (Zhang, Kuila, et al. 2015) and $CoMoO_4$ (Yu, Lu, et al. 2014). SCs based on such hybrid electrodes have shown improved electrochemical performance. Within these composites, large pseudocapacitance is contributed by metal oxides due to rich surface/near-surface redox reactions and graphene contribute to an electric dual-layer capacitance through the adsorption/desorption of electrolyte ions. Simultaneously, metal oxide nanoparticles can be supported by the robust framework of graphene, which provides a conductive network for convenient diffusion of electrolyte ions and fast transport of electrons throughout the electrode.

**4.1.3** *TMOs/other carbon materials*

Carbon is wildly employed as an electrode material for SCs. In addition to CNTs and graphene, a variety of carbon materials fabricated from a series of carbonization and activation processes such as activated carbon, templated porous carbon, carbide-derived carbon, and onion-like carbon have been used for EDLCs. Each kind of carbon materials possesses its own pros and cons. For example, activated carbons, carbide-derived carbons and templated carbons have high surface areas and rich pore structures, while onion-like carbons show good accessibility for electrolyte ions. In this part, we focus on the porous carbon/TMO hybrid for SCs due to their high surface areas and good electrical conductivity. These porous carbons are always derived by polymer, carbide, metal-organic framework, or biomass, etc. For instance, $MnO_2$ nanocrystals decorated mesoporous carbon (MC) have been synthesized by the redox reaction between permanganate ions and carbons (Dong et al. 2006). Controlling the concentration of permanganate ion could successfully retain the ordered mesoporous structures for carbon material. The respectable electrochemical stability and reversibility can be obtained for this $MnO_2$/MC hybrid with a specific capacitance of 200 F $g^{-1}$. Liu et al. (Liu, Zhang, et al. 2018) fabricated hierarchical NiO/C hollow sphere composite by using $SiO_2$ as a hard template and dopamine as a carbon source (**Figure 21a**). The hot alkaline environment during the formation of the $Ni(OH)_2$ nanosheets can etch the $SiO_2$ template without further removal. The hollow carbon core reduced the possibility of ion depletion, while the thin layer of NiO effectively shortened the ion transfer pathway. The well-dispersed hollow structure delivered competitive electrochemical performance.

Ordered mesoporous carbon (OMC) materials are recognized as an ideal host because of their chemical inertness, uniform pore size, and biocompatibility (Zhi et al. 2015). The mesoporous channels (4-6 nm) in OMC induce a low ion-transport resistance. Importantly, the nucleation and growth of metal oxide nanocrystals in carbon mesopores is a confined process, so the control of the nanoparticle growth to obtain reasonable dimension and morphology is feasible. Until now, various nanoparticles decorative OMC composite have been studied by different methods, such as ion-exchange methods, impregnation, sonochemical methods, etc. (Yang et al. 2002; Zhu et al. 2005). One of the representative examples was reported by Zhou et al. (Zhu et al. 2005), they deposited $MnO_2$ nanoparticles inside the pore channels of OMC (denoted as CMK-3) by a sonochemical method. The proposed process and the microstructure of the materials can be found in **Figure 21b,c**. The $MnO_2$ nano-particles were uniformly dispersed in nanochannels with uniform sizes of ∼3-4 nm, and no bulk aggregation of nanoparticles was observed on the outer surface. The specific surface area of CMK-3-$MnO_2$ was 700-900 $m^2$ $g^{-1}$, comparable to single CMK-3 (968 $m^2$ $g^{-1}$). The composite showed a high capacitance of 605 F $g^{-1}$ for $MnO_2$ at a scan rate of 5 mV $s^{-1}$ in the aqueous $Na_2SO_4$ electrolyte. Another example is provided by Hu et al. (Hu, Noked, et al. 2015), where they developed a double-template method fused both hard template and soft template by anodized aluminum oxide (AAO) and triblock copolymer F127 for the formation of ordered mesoporous carbon nanowires (OMCNWs), which can be used as host material for $Fe_2O_3$ nanoparticles (**Figure 21d,e**). The OMCNW/$Fe_2O_3$ composite not only shows large pore volume and surface area but also maintains its structural stability even loading a high amount of metal oxide. Furthermore, the composite demonstrates rapid ion mobility due to the unique nanowire morphology and mesoporous structure, enabling excellent electrochemical performance with high capacitance, good rate capability, and cycling stability.

Carbon coatings, especially core-shell nanostructures, have been designed as protective and electrically conductive layers to improve the stability and conductivity of TMO micro/nanostructures (Shin et al. 2017; Shin et al. 2018). Zheng' group (Zhang, Zhang, et al. 2016) synthesized graphitic carbon coated hollow CuO sphere hybrid through the carbonization of glucose on the hollow CuO sphere (**Figure 21f**). The hybrid showed a high specific surface area (106.6 $m^2$ $g^{-1}$), penetrated mesochannels (~5-15 nm), a large pore volume (0.313 cm3 $g^{-1}$), a robust hollow structure, and an integral graphitic carbon layer. The ACS cell coupled with active carbon exhibited high energy density of 38.6 W h $kg^{-1}$ at a power density of 1.018 kW $kg^{-1}$.

Apart from the ex-situ synthesized carbon as described above, some in-situ methods have been attracted much attention owing to their simple synthesis process. Recently, metal-organic frameworks (MOFs) derived carbons have been reported to show controllable porous architectures, pore volumes, and surface areas. Hence MOFs are considered as suitable precursors for the derivation of porous TMO-carbon composites. For example, a graphitic carbon/chromium oxide ($Cr_2O_3$/C) nanoribbon composite (**Figure 21g**) was prepared by carbonization of a mixture of polyfurfuryl alcohol and MIL-101 under Ar flow at 900 °C (Ullah et al. 2015). Besides the carbon and $Cr_2O_3$ sources, MIL-101(Cr) acted as a template to accommodate furfuryl alcohol as the primary carbon source. The $Cr_2O_3$/C composite with enhanced electronic conductivity and surface area showed a high capacitance (291 F $g^{-1}$) and kept stable (95.5% after 3000 cycles). Other MOF-derived materials such as porous hollow $Co_3O_4$/N-C polyhedron (Kang et al. 2017), ternary $MoO_2$@Cu@C composites (Zhang, Lin, Sun, et al. 2016), and $Co_3O_4$/nanoporous carbon (Young et al. 2018) have been studied to show outstanding electrochemical performance.

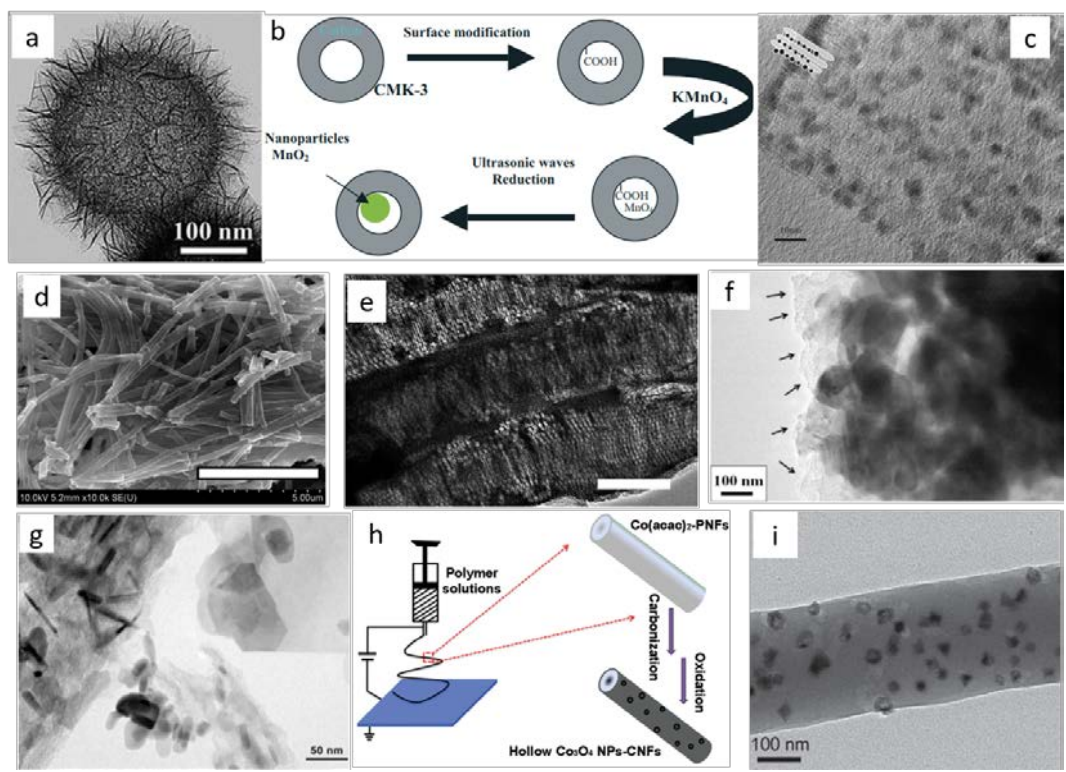

**Figure 21.** (a) TEM image of NiO/C. Reproduced with permission (Liu, Zhang, et al. 2018). Copyright 2018, Royal of Society of Chemistry. (b) The proposed formation process of $MnO_2$ nanoparticles inside the pores of CMK-3, and (c) CMK-$MnO_2$ structural model of the side view of the pore walls with nanocrystals formed along the pore channels of CMK-3. Reproduced with permission (Zhu et al. 2005). Copyright 2005, Wiley. (d) SEM and (e) TEM image of OMCNW/$Fe_2O_3$, scale bar in (e) is 200 nm. Reproduced with permission (Hu, Noked, et al. 2015). Copyright 2015, Royal of Society of Chemistry. (f) TEM image of mesoporous hollow CuO@C spheres. Reproduced with permission (Zhang, Zhang, et al. 2016). Copyright 2016, American Chemical Society. (g) TEM image of $Cr_2O_3$/C nanocomposite. Reproduced with permission (Ullah et al. 2015). Copyright 2015, Elsevier. (h) Schematic illustration of the preparation process of the hollow $Co_3O_4$ NPs-CNFs via an electrospinning and post annealing strategy. (i) TEM image of $Co_3O_4$ NPs-CNFs. Reproduced with permission (Zhang, Yuan, et al. 2013). Copyright 2013, Wiley.

One-dimensional carbon nanofibers (CNFs) based composite have been another promising candidate of electrode material for SCs due to their high conductivities, good mechanical integrity, and large surface area (Wang, Yang, et al. 2011). These materials are always synthesized by electrospinning combined with in-situ calcination, which is a simple, low cost, and scalable technology with tailored nanostructures and compositions. The electrospun CNF/MO composite can be directly used as a free-standing electrode with flexible features. One of the typical examples was reported by

Zhang and Lou (Zhang, Yuan, et al. 2013), they obtained $Co_3O_4$ hollow nanoparticles (NPs) embedded CNFs by electrospun (**Figure 21h**). The $Co_3O_4$ hollow NPs were uniformly located in the CNF (**Figure 21i**). The flexible mat-like films with tunable $Co_3O_4$ loadings showed advantageous structural features, resulting in a high SC of 556 F $g^{-1}$ at a current density of 1 A $g^{-1}$, and 403 F $g^{-1}$ even at a very high current density of 12 A $g^{-1}$.

**4.2** TMOs/conducting polymers

Conducting polymers (CPs) are a kind of organic polymers which have conductivity through a conjugated bond matrix in the polymer molecule. CPs have attracted extensive interest for diverse SC applications because of their high energy density, low cost, environmental friendliness, and reversible Faradaic redox capabilities. PANi and PPy are the most developed conducting polymer for SCs. Nowadays, CPs have been found to be a promising candidate for TMO composite due to their good conductivity, flexibility, relatively cheap and easy of synthesis.

PANi is synthesized by polymerization of aniline monomer by various approaches. Many TMOs such as $MnO_2$, CuO, and $NiCoO_4$ have been hybridized with PANi (Jiang et al. 2012; Xu, Wu, et al. 2015; Ates et al. 2015). Jiang et al. (Jiang et al. 2012) deposited ultrafine and loosened K-birnessite $MnO_2$ floccules on the surface of the PANi nanofibers via simply soaking the PANi nanofibers in a $KMnO_4$ aqueous solution (**Figure 22a,b**). This composite showed several merits for supercapacitors: (a) the high specific surface area provides more electroactive sites; (b) one-dimensional nanofibers with high conductivity can serve as both the backbone and conductive pathway for $MnO_2$; (c) Both the $MnO_2$ and PANi are good pseudo-capacitive materials. As shown in **Figure 22c,d**, the composite displayed high specific capacitance of 383 F $g^{-1}$, stable cycling performance, and high energy and power density (53.2 W h $kg^{-1}$ at a power density of 250 W $kg^{-1}$ and 18.6 W h $kg^{-1}$ even at a high power density of 10000 W $kg^{-1}$).

PPy, synthesized by polymerization of pyrrole monomer, have high charge storage ability and high conductivity. The specific capacitance of pure PPy is usually less than 400 F g$^{-1}$. The PPy/TMO composite has been found to possess ultra-high specific capacitance and ultra-stable cycle performance (Wang, Zhan, et al. 2015; Qian, Zhou, et al. 2015; Sun, Li, et al. 2015; Zhou et al. 2013). For instance, the capacitance of WO$_3$/PPy core/shell nanowires can be achieved 253 mF cm$^{-2}$ (Wang, Zhan, et al. 2015). When assembling the ASC cell with carbon fibers/Co(OH)$_2$ electrode, it exhibited high volumetric capacitance and energy density (2.865 F cm$^{-3}$ and 1.02 mWh cm$^{-3}$), as well as high cycling stability (~90.5% of retention after 4000 cycles). Li and Liu (Zhou et al. 2013) designed a novel hybrid electrode which was formed by the growth of well-aligned CoO nanowire array on 3D nickel foam and then homogenously coated or firmly anchored the PPy onto the surface of nanowires (**Figure 22e-g**). This novel architecture showed merits of the high electrochemical activity of CoO, the high electronic conductivity of PPy, and the short ion diffusion pathway in ordered mesoporous nanowires. The synergetic effect between CoO and PPy induced outstanding performance, including a high specific capacitance of 2223 F g$^{-1}$ which is close to the theoretical value, good rate capability, and cycling stability (99.8% capacitance retention after 2000 cycles) (**Figure 22h**).

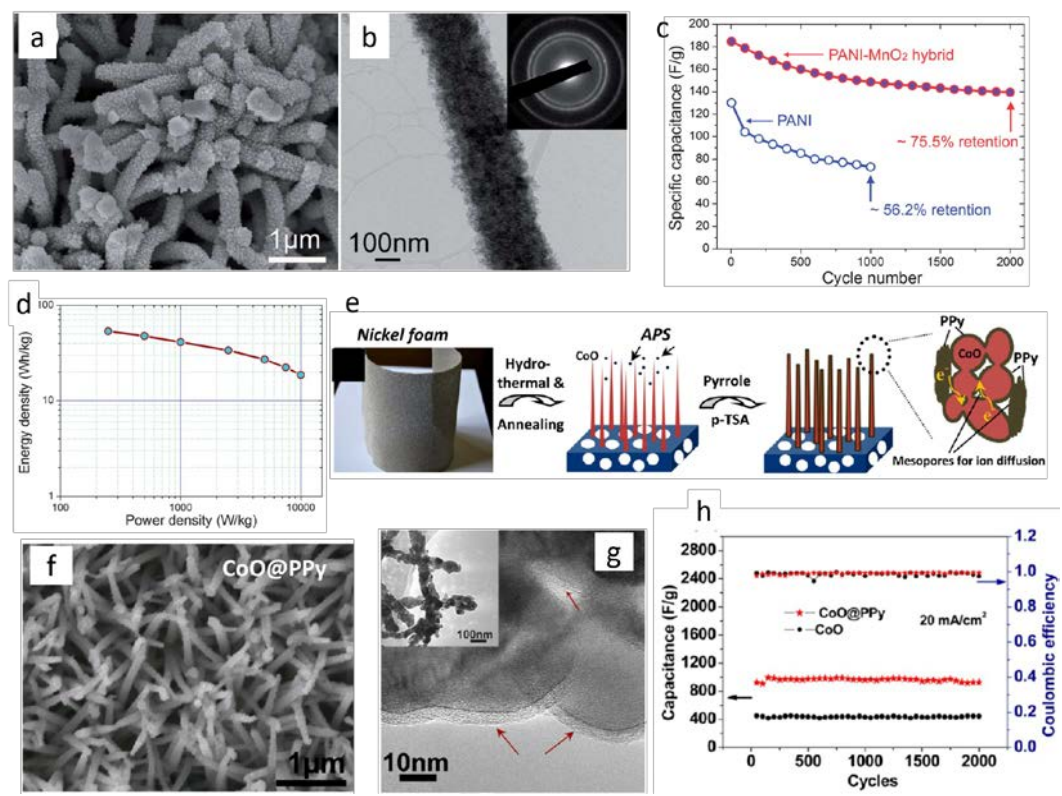

**Figure 22.** (a) SEM image and (b) TEM image (inset showing the SAED pattern) the PANI–MnO$_2$ coaxial nanofibers. (c) Specific capacitance as a function of cycle number at 10 A g$^{-1}$ of the PANI–MnO$_2$ coaxial nanofibers. (d) Ragone plot of the optimal PANI–MnO$_2$ coaxial nanofibers. Reproduced with permission (Wang, Zhan, et al. 2015). Copyright 2015, Wiley. (e) The representative synthetic procedure and structure details of the 3D CoO@PPy hybrid nanowire electrode. (f) SEM images of the hybrid nanowire electrode. (g) HRTEM images of the surface of individual CoO@PPy hybrid nanowires. Inset shows the general view of several nanowires. (h) Cycling performance of CoO@PPy hybrid electrode and the pristine CoO electrode. Reproduced with permission (Zhou et al. 2013). Copyright 2013, American Chemical Society.

### 4.3 TMOs/metal compounds composites

The hierarchical structure composed by TMO and the metal-based compound is a promising solution for high energy density and high power density SCs because of the synergetic effect of the two components both of which possess pseudocapacitance behavior with high capacitance. The main synthetic guidance method for the hierarchical structure includes (a) design of the hierarchical structure based on transition metal material powder; (b) building composite materials using binder-free array structure as the backbone for the growth of another material; (c) self-assembly of the same transition metal compounds into hierarchical structures. The SC performance has been reported to be significantly improved by constructing hierarchical structures for TMOs with metal, metal oxides/hydroxides, sulfides, selenides, and phosphides. For example, Qu et al. (Qu et al. 2013) synthesized Au-decorated hierarchical NiO nanostructures (Au-NiO) through presented a simple solution method. When the Au-NiO composite was used for pseudocapacitors, not only a remarkable cycling stability of 690 F $g^{-1}$ at 4 A $g^{-1}$ after 2000 cycles could be achieved, but also an excellent rate capability of 619 F $g^{-1}$ at 20 A $g^{-1}$ was reached. Ternary ordered hierarchical $Co_3O_4$@Ni-Co-O nanosheet-nanorod arrays show high efficiency of 2098 F $g^{-1}$, giving specific capacitances per area as high as ∼25 F $cm^{-2}$ at 5 mA $cm^{-2}$ with a high mass loading of 12 mg $cm^{-2}$ (Lu, Yang, et al. 2012).

The TMO/TMO composite-based hierarchically porous or hollow structures have been developed for SCs (Wei et al. 2014; Tan et al. 2016; Zhou et al. 2018). For instance, Wei et al. (Wei et al. 2014) developed 3D ZnO-NiO mesoporous architectures as electrochemical capacitors. The 3D ZnO-NiO composite calcinated at 400 °C exhibited a high specific capacitance of 2498 F $g^{-1}$ at 2.6 A $g^{-1}$ by the electrochemical tests. Additionally, this composite also showed good rate capability at high current densities and excellent long-term cycling stability (about 3.0% loss of the maximum specific capacitance after 2000 cycles), which was attributed to the morphological features of mesoporous and nanosheet self-assembled architecture, as well as a rational composition of the two constituents. Many TMO/TMO composite, such as $MoO_3$@CuO (Zhang, Lin, Wang, et al. 2016), $Co_3O_4$/$NiCo_2O_4$ (Yu, Wu, et al. 2016), ZnO/NiO (Zhang, Zhang, et al. 2017), $Co_3O_4$/$ZnFe_2O_4$ (Hu, Liu, et al. 2015), ZnO@C@$NiCo_2O_4$ (Zeng, Wang, et al. 2016), have been successfully derived by MOFs (POMs@MOFs, ZIF-67, Heterobimetallic, Fe(III)-MOF-5, ZIF-8).

Especially, core-shell structures conducted by TMO-based composite have been reported to possess advanced contributions for SCs. CuO@$MnO_2$ core-shell nanostructures were synthesized by a simple and cost-effective method without any surfactants (Huang et al. 2014). The complex hetero-structures demonstrated a high specific capacitance of 276 F $g^{-1}$ at a current density of 0.6 A $g^{-1}$, much enhanced rate performance, and long-term cycling stability (92.1% retention after 1000 cycles).

Recently, the hierarchical structures have been designed on the base of electroactive arrays (Yu and Thomas 2014; Liu et al. 2011; Xia, Chao, et al. 2014). For example. Mai at al. (Mai et al. 2011) successfully synthesized a 3D multicomponent oxide hierarchical heterostructures ($MnMoO_4/CoMoO_4$), which were prepared by surface modification on the backbone material $MnMoO_4$ through a simple refluxing method under mild conditions. The crystal growth mechanism during the complicated nano-architecture process is "self-assembly" and "oriented attachment". The asymmetric SCs fabricated by hierarchical $MnMoO_4/CoMoO_4$ heterostructured nanowires showed a specific capacitance of 187.1 F g$^{-1}$ at a current density of 1 A g$^{-1}$, and good reversibility with a cycling efficiency of 98% after 1000 cycles, which are much higher than that for either pure 1D $MnMoO_4$ nanorods or $CoMoO_4$ nanowires.

Among the various electroactive arrays, Ni-Co sulfide arrays have been considered as a promising candidate due to the metallic electrical conductivity and high electrochemical activity. As shown in **Figure 23**, Xiao et al. (Xiao et al. 2014) grew highly conductive $NiCo_2S_4$ single crystalline nanotube arrays on a flexible carbon fiber paper (CFP). The good pseudocapacitive material $NiCo_2S_4$ served as a 3D conductive scaffold for loading additional electroactive metal oxide materials, including $Co_xNi_{1-x}(OH)_2$, $MnO_2$, and FeOOH. $NiCo_2S_4$ showed much higher electrical conductivity than $NiCo_2O_4$ nanorod arrays, resulting in superior electrochemical performance for $NiCo_2S_4$. After deposition of three materials, the as-prepared $Co_xNi_{1-x}(OH)_2$/ $NiCo_2S_4$ nanotube array electrodes showed the highest discharge areal capacitance (2.86 F cm$^{-2}$ at 4 mA cm$^{-2}$), good rate capability (still 2.41 F cm$^{-2}$ at 20 mA cm$^{-2}$), and excellent cycling stability (~4% loss after the repetitive 2000 cycles at a charge-discharge current density of 10 mA cm$^{-2}$).

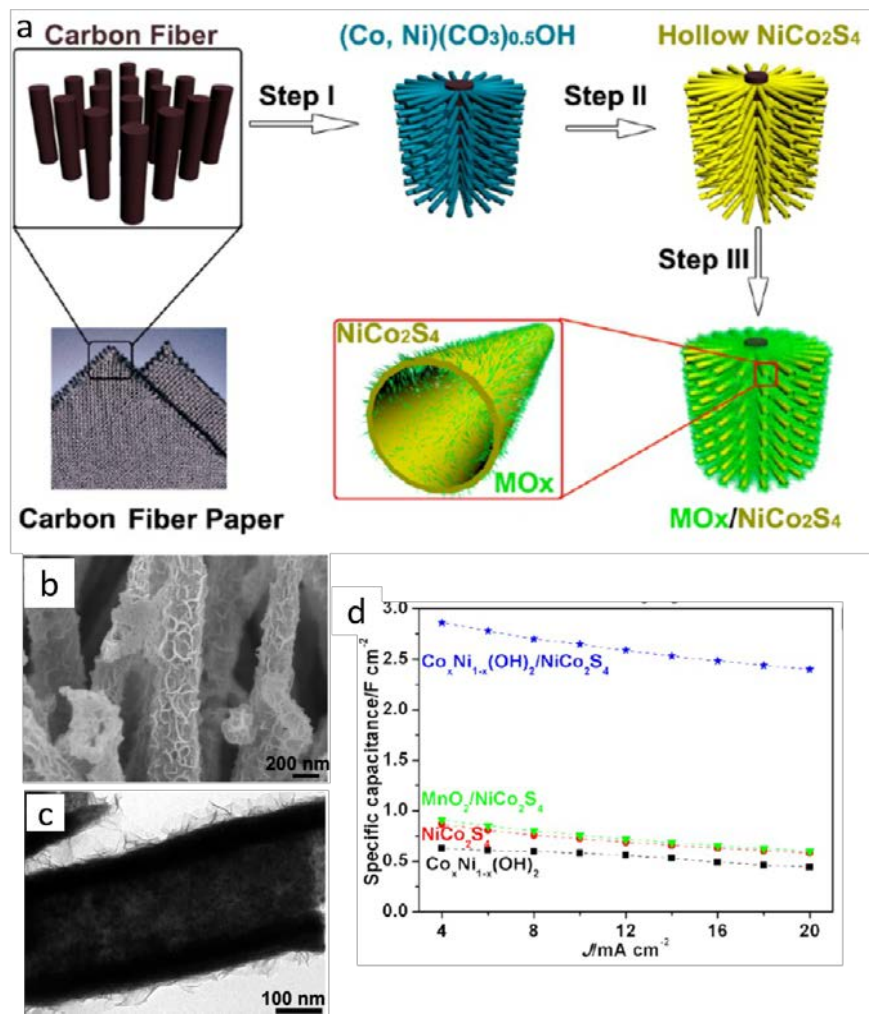

**Figure 23.** (a) Nanofabrication flowchart of the hierarchically structured composite electrodes of electroactive materials ($MO_x$)/$NiCo_2S_4$. (b) SEM and (c) TEM images of the $Co_xNi_{1-x}(OH)_2$/$NiCo_2S_4$. (d) The discharge areal capacitance performance of different electrodes. Reproduced with permission (Xiao et al. 2014). Copyright 2014, American Chemical Society.

## 5. Conclusions and outlook

Metal oxide-based supercapacitive electrodes are promising to dramatically promote the capacitance of SCs for their faradic charge-storage process. MO electrodes for SCs can be classified into three types based on the charge storage mechanisms: surface redox pseudocapacitance based materials, intercalation pseudocapacitance based materials, and battery-type materials. All of the MO materials deliver much higher energy density than carbon-based materials with electric double layer capacitance. However, the issues for MO-based electrode are the relatively low conductivity and slow ion transfer, which probably lead to deterioration in the rate capability and cycle stability. The electrochemical performance of MO electrode is highly influenced by their crystal structures, electric conductivity, and charge storage mechanism. To address this concern, nanostructure design, an effective approach for the nanomaterials involved interface reaction, can demonstrate a large surface area, shorten the diffusion paths of electrolyte ions and electrons, and supply abundant excess interspace for buffering the big volume expansion of active materials during charge/discharge processes. In addition, hybrid nanostructures composed of metal oxide and a conducting material matrix such as carbon materials (CNT, graphene, porous carbon, coating carbon, carbon nanofiber) and conductive polymers, or combined with other metal-based compounds, should be finely established, which can enlarge the surface area, improve the electrical conductivity, enhance the morphological stability, and build enhanced synergies as well, which leads to improved capacitance, rate capability and longer cycling stability.

Certain challenges remain to be overcome and considered for fabricating high-performance and cost-effective electrochemical energy storage device. Advanced construction and optimization of material and electrode structure need to be highlighted and investigated. For instance, the properties of interfaces of electrode active materials/electrolyte and of electrode active materials/substrates are critical to the electrochemical kinetics and charge collection of SCs. Engineering interfaces to reduce the barrier for charge transport and possible unwanted reactions have attracted much attention. Some strategies including element doping, the introduction of vacancies, and surface functionalization have been initially developed but not satisfactory. Meanwhile, the technologies for the enhancement of current collect/electrode interfaces such as binder-free electrode, buffer layer formation, and novel current collector development are also needed to be explored in advance. More importantly, the electrochemical reaction mechanisms for many metal oxides in various supercapacitor system are still unclear and worthy of deep investigation. Advanced electrochemical examination and material characterization techniques may be needed. The weight loading and the thickness of the electrode should be addressed for the device application. Moreover, the electrolyte and separator applied for the full device also need to be taken into consideration.